\documentclass[twocolumn]{aastex62}
\usepackage[utf8]{inputenc}
\DeclareUnicodeCharacter{2212}{-}
\usepackage{graphicx}
\graphicspath{{./}{Figures}}
\usepackage{amsmath}
\usepackage{afterpage}
\usepackage{enumitem}
\usepackage{xcolor}
\usepackage{float}
\usepackage{xspace} 
\usepackage{url}
\usepackage{fontawesome}
\setenumerate{itemsep=0mm}
\usepackage{wrapfig}
\usepackage{placeins}

\usepackage{amsmath}
\usepackage{amssymb}
\usepackage{xspace}
\usepackage{xifthen}
\usepackage{eso-pic}

\newcommand{\Gaia}{{\it Gaia}\xspace}

\definecolor{forestgreen}{HTML}{228B22}
\definecolor{urlblue}{HTML}{000000}

\mathchardef\mhyphen="2D

\newlength{\dhatheight}

\newcommand{\code}[1]{\texttt{#1}\xspace}

\newcommand{\unit}[1]{\ensuremath{\mathrm{\,#1}}\xspace}

\newcommand{\Gyr}{\unit{Gyr}}

\newcommand{\degree}{\ensuremath{{}^{\circ}}\xspace}

\newcommand{\secref}[1]{Section~\ref{sec:#1}}

\newcommand{\tabref}[1]{Table~\ref{tab:#1}}

\newcommand{\figref}[1]{Figure~\ref{fig:#1}}

\newcommand{\bandvar}[2][]{%
  \ifthenelse{\isempty{#1}}{\var{#2}}{\var{#2\_#1}}%
}

\newcommand{\ra}{{\ensuremath{\alpha_{2000}}}\xspace}
\newcommand{\dec}{{\ensuremath{\delta_{2000}}}\xspace}
\newcommand{\age}{{\ensuremath{\tau}}\xspace}

\newcommand{\feh}{{\ensuremath{\rm [Fe/H]}}\xspace}
\newcommand{\ellip}{\ensuremath{\epsilon}\xspace}
\newcommand{\PA}{\ensuremath{\mathrm{P.A.}}\xspace}

\newcommand{\HEALPix}{\code{HEALPix}}
\newcommand{\healpix}{\HEALPix}

\newcommand{\emcee}{\code{emcee}}
\newcommand{\ugali}{\code{ugali}}

\newcommand{\var}[1]{\ensuremath{\texttt{\MakeUppercase{#1}}}\xspace}

\providecommand\physrep{\ref@jnl{Phys.~Rep.}}%
\providecommand\apjs{\ref@jnl{ApJS}}%
\providecommand{\jcap}{\ref@jnl{JCAP}}%

\reportnum{\footnotesize FERMILAB-PUB-23-271-LDRD-PPD}

\shorttitle{An Ancient, Ultra-Faint Star Cluster Near the Magellanic Clouds}

\shortauthors{DELVE Collaboration}

\begin{document}

\reportnum{\footnotesize}

\title{DELVE 6: An Ancient, Ultra-Faint Star Cluster on the Outskirts of the Magellanic Clouds}


\correspondingauthor{William Cerny}
\email{william.cerny@yale.edu}

\author[0000-0003-1697-7062]{W.~Cerny}
\affiliation{Department of Astronomy, Yale University, New Haven, CT 06520, USA}

\author[0000-0001-8251-933X]{A.~Drlica-Wagner}
\affiliation{Fermi National Accelerator Laboratory, P.O.\ Box 500, Batavia, IL 60510, USA}
\affiliation{Kavli Institute for Cosmological Physics, University of Chicago, Chicago, IL 60637, USA}
\affiliation{Department of Astronomy and Astrophysics, University of Chicago, Chicago IL 60637, USA}

\author[0000-0002-9110-6163]{T.~S.~Li}
\affiliation{Department of Astronomy and Astrophysics, University of Toronto, 50 St. George Street, Toronto ON, M5S 3H4, Canada}

\author[0000-0002-6021-8760]{A.~B.~Pace}
\affiliation{McWilliams Center for Cosmology, Carnegie Mellon University, 5000 Forbes Ave, Pittsburgh, PA 15213, USA}

\author[0000-0002-7134-8296]{K.~A.~G.~Olsen}
\affiliation{NSF's National Optical-Infrared Astronomy Research Laboratory, 950 N. Cherry Ave., Tucson, AZ 85719, USA}

\author{N.~E.~D.~No\"el}
\affiliation{Department of Physics, University of Surrey, Guildford GU2 7XH, UK}

\author[0000-0001-7827-7825]{R.~P.~van~der~Marel}
\affiliation{Space Telescope Science Institute, 3700 San Martin Drive, Baltimore, MD 21218, USA}
\affiliation{Center for Astrophysical Sciences, Department of Physics \& Astronomy, Johns Hopkins University, Baltimore, MD 21218, USA}

\author[0000-0002-3936-9628]{J.~L.~Carlin}
\affiliation{Rubin Observatory/AURA, 950 North Cherry Avenue, Tucson, AZ, 85719, USA}

\author[0000-0003-1680-1884]{Y.~Choi}
\affiliation{Department of Astronomy, University of California, Berkeley, Berkeley, CA 94720, USA}

\author[0000-0002-8448-5505]{D.~Erkal}
\affiliation{Department of Physics, University of Surrey, Guildford GU2 7XH, UK}

\author[0000-0003-1697-7062]{M.~Geha}
\affiliation{Department of Astronomy, Yale University, New Haven, CT 06520, USA}

\author[0000-0001-5160-4486]{D.~J.~James}
\affiliation{ASTRAVEO LLC, PO Box 1668, Gloucester, MA 01931}
\affiliation{Applied Materials Inc., 35 Dory Road, Gloucester, MA 01930}

\author[0000-0002-9144-7726]{C.~E.~Mart\'inez-V\'azquez}
\affiliation{Gemini Observatory, NSF's NOIRLab, 670 N. A'ohoku Place, Hilo, HI 96720, USA}

\author[0000-0002-8093-7471]{P.~Massana}
\affiliation{Department of Physics, Montana State University, P.O. Box 173840, Bozeman, MT 59717-3840}

\author[0000-0003-0105-9576]{G.~E.~Medina}
\affiliation{Department of Astronomy and Astrophysics, University of Toronto, 50 St. George Street, Toronto ON, M5S 3H4, Canada}

\author[0000-0002-7483-7327]{A.~E.~Miller}
\affiliation{School of Mathematical and Physical Sciences, Macquarie University, Balaclava Road, Sydney, NSW 2109, Australia}
\affiliation{Research Centre for Astronomy, Astrophysics and Astrophotonics, Macquarie University, Balaclava Road, Sydney, NSW 2109, Australia}
\affiliation{Leibniz-Institut f\"{u}r Astrophysik Potsdam (AIP), An der Sternwarte 16, D-14482 Potsdam, Germany}
\affiliation{Institut f\"{u}r Physik und Astronomie, Universit\"{a}t Potsdam, Haus 28, Karl-Liebknecht-Str. 24/25, D-14476 Golm (Potsdam), Germany}

\author[0000-0001-9649-4815]{B.~Mutlu-Pakdil}
\affiliation{Department of Physics and Astronomy, Dartmouth College, Hanover, NH 03755, USA}

\author{D.~L.~Nidever}
\affiliation{Department of Physics, Montana State University, P.O. Box 173840, Bozeman, MT 59717-3840; NSF's National Optical-Infrared Astronomy Research Laboratory, 950 N. Cherry Ave., Tucson, AZ 85719, USA}
\affiliation{NSF's National Optical-Infrared Astronomy Research Laboratory, 950 N. Cherry Ave., Tucson, AZ 85719, USA}

\author[0000-0002-1594-1466]{J.~D.~Sakowska}
\affiliation{Department of Physics, University of Surrey, Guildford GU2 7XH, UK}

\author[0000-0003-1479-3059]{G.~S.~Stringfellow}
\affiliation{Center for Astrophysics and Space Astronomy, University of Colorado, 389 UCB, Boulder, CO 80309-0389, USA}

\author[0000-0002-3690-105X]{J.~A.~Carballo-Bello}
\affiliation{Instituto de Alta Investigaci\'on, Sede Esmeralda, Universidad de Tarapac\'a, Av. Luis Emilio Recabarren 2477, Iquique, Chile}

\author[0000-0001-6957-1627]{P.~S.~Ferguson}
\affiliation{Department of Physics, University of Wisconsin-Madison, Madison, WI 53706, USA}

\author[0000-0003-2511-0946]{N.~Kuropatkin}
\affiliation{Fermi National Accelerator Laboratory, P.O.\ Box 500, Batavia, IL 60510, USA}

\author[0000-0003-3519-4004]{S.~Mau}
\affiliation{Department of Physics, Stanford University, 382 Via Pueblo Mall, Stanford, CA 94305, USA}
\affiliation{Kavli Institute for Particle Astrophysics \& Cosmology, P.O.\ Box 2450, Stanford University, Stanford, CA 94305, USA}

\author[0000-0002-9599-310X]{E.~J.~Tollerud}
\affiliation{Space Telescope Science Institute, 3700 San Martin Drive, Baltimore, MD 21218, USA}

\author[0000-0003-4341-6172]{A.~K.~Vivas}
\affiliation{Cerro Tololo Inter-American Observatory, NSF's National Optical-Infrared Astronomy Research Laboratory,\\ Casilla 603, La Serena, Chile}

\collaboration{(DELVE Collaboration)}

\begin{abstract}

We present the discovery of DELVE 6, an ultra-faint stellar system identified in the second data release of the DECam Local Volume Exploration (DELVE) survey. Based on a maximum-likelihood fit to its structure and stellar population, we find that DELVE 6 is an old ($\tau > 9.8$~Gyr, at 95\% confidence) and metal-poor ($\rm [Fe/H] < -1.17$~dex, at 95\% confidence) stellar system with an absolute magnitude of $M_V = -1.5^{+0.4}_{-0.6}$~mag and an azimuthally-averaged half-light radius of $r_{1/2} =10^{+4}_{-3}$~pc. These properties are consistent with the population of ultra-faint star clusters uncovered by recent surveys.  Interestingly, DELVE 6 is located at an angular separation of $\sim 10\degree$ from the center of the Small Magellanic Cloud (SMC), corresponding to a three-dimensional physical separation of $\sim 20$~kpc given the system's observed distance ($D_{\odot} = 80$~kpc). This also places the system $\sim 35$~kpc from the center of the Large Magellanic Cloud (LMC), lying within recent constraints on the size of the LMC's dark matter halo. We tentatively measure the proper motion of DELVE 6 using data from \textit{Gaia}, which we find supports a potential association between the system and the LMC/SMC. Although future kinematic measurements will be necessary to determine its origins, we highlight that DELVE~6 may represent only the second or third ancient ($\tau > 9$~Gyr) star cluster associated with the SMC, or one of fewer than two dozen ancient clusters associated with the LMC. Nonetheless, we cannot currently rule out the possibility that the system is a distant Milky Way halo star cluster.
\end{abstract}



\section{Introduction}
\label{sec:intro}
Recent large-scale digital sky surveys have revolutionized our understanding of the Magellanic Clouds (MCs) and their environments. In particular, sensitive surveys with the VISual and Infrared Telescope for Astronomy, (e.g., VMC; \citealt{2011A&A...527A.116C}), the VLT Survey Telescope (e.g., STEP and YMCA; \citealt{2014MNRAS.442.1897R}; \citealt{2021RNAAS...5..159G}), and the Dark Energy Camera on the 4m Blanco Telescope (e.g., DES, SMASH, and MagLiteS; \citealt{2005astro.ph.10346T}; \citealt{2017AJ....154..199N}; \citealt{2017AAS...22941606B}) have provided an unprecedentedly deep view of the diverse stellar populations of the MCs, enabling detailed characterization of their star formation histories \citep[e.g.,][]{2015MNRAS.449..639R,2018MNRAS.478.5017R,2021MNRAS.508..245M, 2022MNRAS.513L..40M}, 3D geometries \citep[e.g.,][]{2017MNRAS.472..808R,2018ApJ...866...90C,2022MNRAS.512..563R},  and substructures \citep[e.g.,][]{2017MNRAS.468.1349P,2018ApJ...858L..21M, 2018ApJ...869..125C,2020MNRAS.498.1034M,2022ApJ...931...19G,2021MNRAS.505.2020E,2022MNRAS.510..445C}. Furthermore, these surveys have significantly expanded the census of star clusters and satellite galaxies in the main bodies and outskirts of the Clouds \citep[e.g.,][]{Bechtol:2015,2015ApJ...805..130K, 2016ApJ...830L..10M, 2016ApJ...833L...5D,2018MNRAS.479.5343K,2018MNRAS.475.5085T, 2021ApJ...910...18C, 2021RNAAS...5..159G}, allowing for constraints on the MCs' masses, dark matter halos, orbits, and interaction histories \citep[e.g.,][]{2016MNRAS.461.2212J, 2018ApJ...867...19K, 2018ApJ...853..104B, 2020MNRAS.495.2554E, 2020ApJ...893..121P,2021A&A...647L...9D} especially when paired with the precise phase-space information provided by the \Gaia satellite \citep{2016A&A...595A...1G, 2022A&A...657A..54B, 2022ApJ...940..136P}. Lastly, these efforts to survey and characterize the satellite populations of the MCs have enabled novel observational tests of hierarchical galaxy formation within the $\Lambda$CDM paradigm at a lower host mass scale than offered by the Milky Way \citep[e.g.,][]{10.1093/mnras/stw2816,  2017MNRAS.472.1060D,10.1093/mnras/stz2457}.

\begin{figure*}
    \centering
     \includegraphics[width = .335\textwidth]{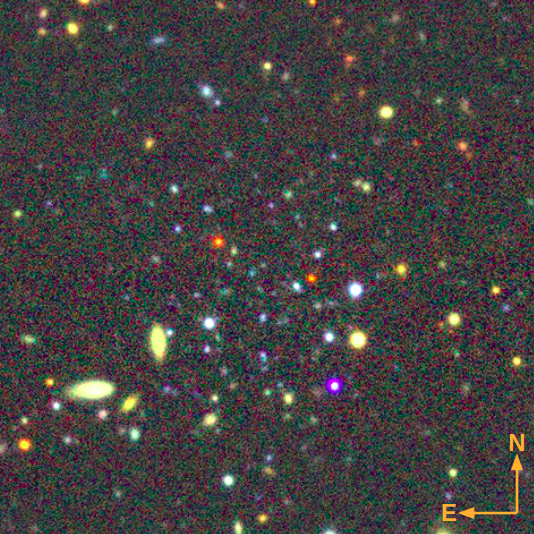}
       \raisebox{-0.12\height}{\includegraphics[width = .425\textwidth]{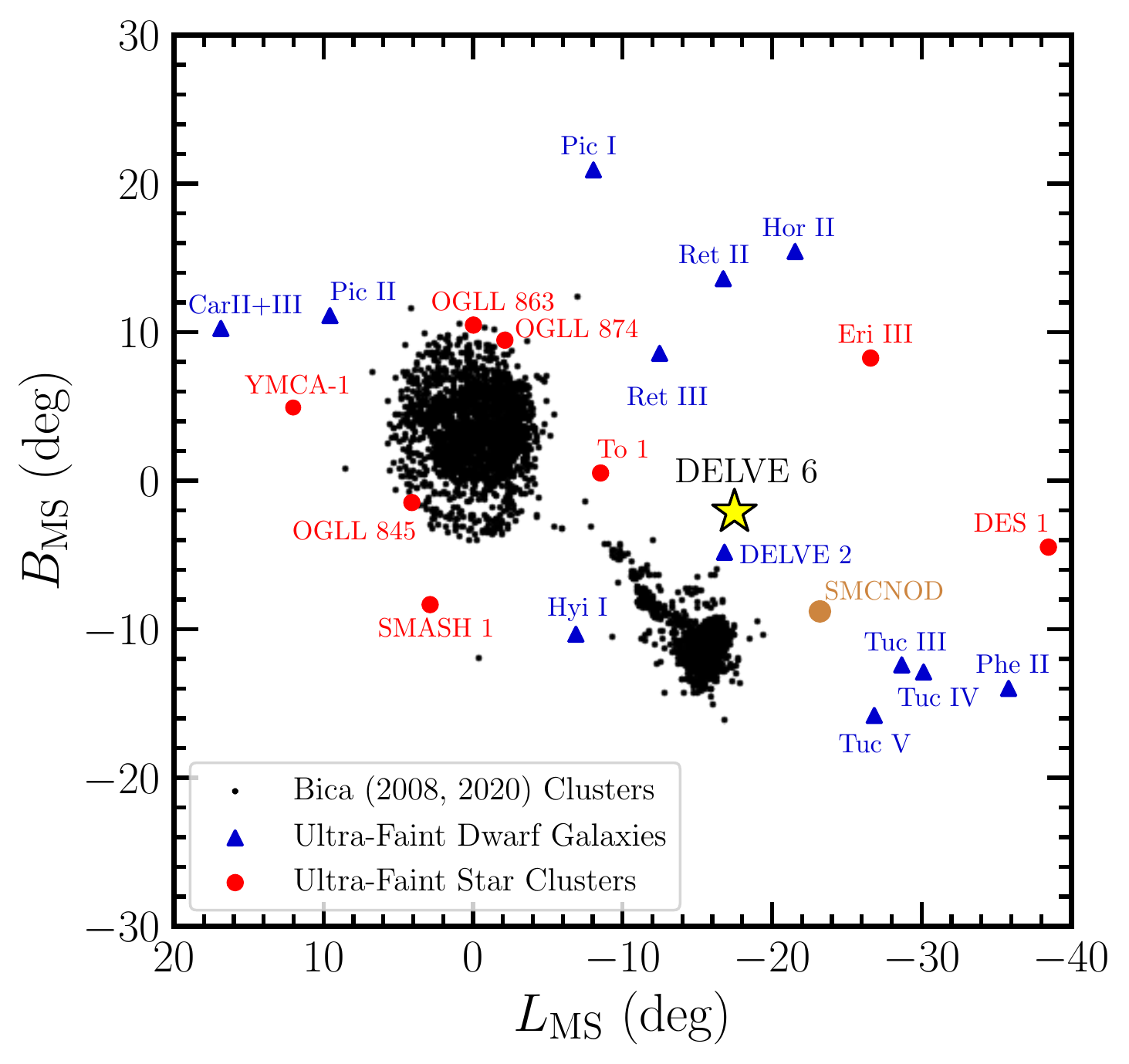}}
    \vspace{-.25em}
    \caption{(Left) $2\arcmin \times 2\arcmin $ false-color image cutout centered on DELVE~6 based on $griz$ DECam imaging, taken from the Legacy Survey Sky Viewer. A clustering of faint blue sources is visible amidst a number of foreground stars and background galaxies. We have increased the brightness of this cutout image to enhance the visibility of faint cluster member stars. (Right) Map of star clusters and dwarf galaxies in the Magellanic environment, plotted in Magellanic Stream coordinate system from \citet{2008ApJ...679..432N}. Each black point corresponds to a star cluster included in the main cluster catalogs from \citet{2008MNRAS.389..678B, 2020AJ....159...82B}. Recently-discovered ultra-faint star clusters in this region are shown as red dots, whereas candidate and confirmed ultra-faint dwarf galaxies potentially associated with the MCs are shown as blue triangles. We caution that some of these objects have uncertain classifications and/or tentative associations with the MCs. In addition, we plot the centroid position of the SMC Northern Overdensity (SMCNOD; \citealt{smcnod}) in orange. Lastly, DELVE 6 is shown as a yellow star, positioned near $B_{\rm MS} \sim 0$; this latitude falls along the projected track of the Magellanic Stream.}
    \label{fig:cutout}
\end{figure*}

\par In this \textit{Letter}, we present the newest discovery in this ongoing census of Magellanic satellites: DELVE~6, an ancient, ultra-faint star cluster in the distant outskirts of the MCs.
This low-mass system was identified through matched-filter searches over imaging from the Dark Energy Camera (DECam; \citealt{2015AJ....150..150F}) processed as part of the second data release of the DECam Local Volume Exploration survey \citep[DELVE DR2;][]{2022ApJS..261...38D}. We find that it has an old and metal-poor stellar population, joining the less-than-two-dozen ancient globular clusters known in the Magellanic environment, and that it falls at an unusually large separation from its likely hosts. Thus, DELVE~6 potentially represents an exciting and novel window into the stellar populations inhabiting the periphery of the LMC/SMC system. Here, we present an initial characterization of this system's basic properties, and briefly highlight possibilities for its origins that can be tested with deeper imaging and spectroscopic followup.

\section{Discovery and Characterization}

\subsection{Identification in DELVE DR2 and the Legacy Surveys DR10}
In \citet{2022arXiv220912422C}, we presented the results of an extensive search for ultra-faint stellar systems in the Milky Way halo using DECam data processed as part of DELVE DR2 \citep{2022ApJS..261...38D}. Briefly, this search involved applying the open-source \code{simple} search code\footnote{https://github.com/DarkEnergySurvey/simple}, which implements an isochrone matched-filter in color--magnitude space to identify overdensities of resolved stars in the Milky Way halo consistent with an old, metal-poor stellar population. Over the entire DELVE DR2 footprint ($\sim$ 21,000 deg$^2$), this search resulted in $\mathcal{O}(10^4)$ overdensities, six of which we confirmed as \textit{bona fide} ultra-faint stellar systems in \citet{2022arXiv220912422C} on the basis of deeper follow-up imaging.

\par During the late stages of preparation of the aforementioned work, new multi-band co-added images built from the DECam data became available through an early version of the Legacy Surveys Data Release 10.\footnote{https://www.legacysurvey.org/dr10/description/} Motivated by the new availability of these images, we performed a visual inspection of a subset of the initial high-significance ($>5.5\sigma$) satellite candidates that resided in regions where comparable color images were not previously available. The primary goal of this effort was to identify systems that were clearly identifiable as overdensities of blue stars in these images, but may have initially been missed due to their marginal signals seen in their smoothed spatial distribution and observed color--magnitude diagrams (CMDs) generated as part of \code{simple}'s diagnostic plots. One such candidate, DELVE~J0212-6603 (DELVE~6), was identified at high significance ($\sim 5-7\sigma$; comparable to the candidates presented in \citealt{2022arXiv220912422C}) in our multi-pronged search, but was initially passed over during prior inspection due to its sparse CMD. However, as seen in \figref{cutout}, this system is visible as a tight clustering of faint blue sources in the color images provided by the Legacy Surveys DR10 and was easily identified during our visual inspection. 
\par After confirming that the system has not been reported in literature catalogs of star clusters and dwarf galaxies in the environment of the MCs \citep[e.g.,][]{2008MNRAS.389..678B, 2020AJ....159...82B, 2020MNRAS.499.4114G}, we proceeded to characterize the newly-discovered system's structure and stellar population, as described in the following subsection. 
In the absence of timely deeper follow-up imaging, we continued to use the photometric catalogs provided by DELVE DR2 for our analysis. The DELVE DR2 data coincident with DELVE 6 are relatively deep, reaching (extinction-corrected) $\rm S/N=10$ magnitude limits of $g_0 \sim 24.0$~mag and $r_0 \sim 23.8$~mag. 
These limits are roughly 0.5 mag and 0.8 mag deeper than the median DELVE DR2 depth in the $g$ and $r$ bands, respectively (see \citealt{2022ApJS..261...38D} for specific information about this public dataset).  This depth was therefore found to be sufficient to characterize the newly-discovered system despite its low luminosity. 

\par Throughout the analyses described below, we separated stars from galaxies based on the selection $0 \leq \code{EXTENDED\_CLASS\_G} \leq 2$, matching our DECam analyses described in \citet{2022arXiv220912422C}.
This broadly allowed for a higher degree of stellar completeness at the cost of increased galaxy contamination at fainter magnitudes (\citealt{2022ApJS..261...38D}).

\begin{figure*}
    \centering
    \includegraphics[width= \textwidth]{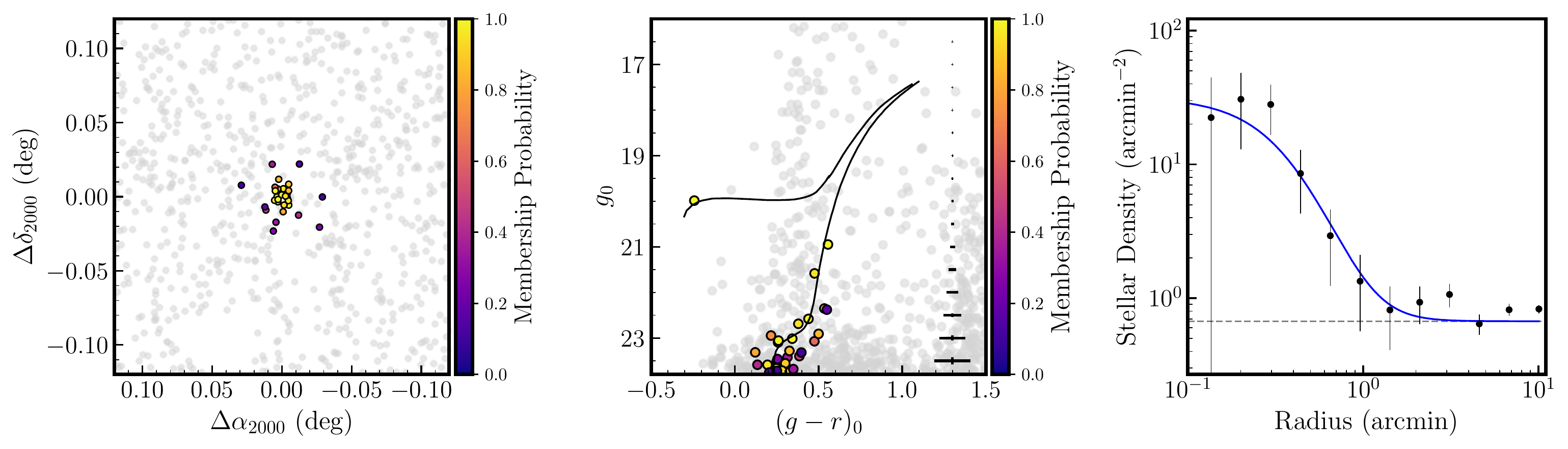}
    \caption{(Left) Spatial distribution of stars in a $0.12\degree \times 0.12\degree$ ($7.2\arcmin \times 7.2\arcmin$) field centered on DELVE~6. Stars are colored by their probability of being a member of the candidate system, as determined through our \ugali fit described in \secref{prop}; stars with probabilities $p < 0.1$ are shown in grey. (Center) CMD for the stars shown in the left panel. The best-fit isochrone with $Z = 0.0001$ and $\tau = 13.5$~Gyr is shown in black, although see the discussion in \secref{prop} and see \figref{amr} for caveats about this model. (Right) Radial stellar density profile for DELVE~6. The best-fit Plummer model is shown as a solid blue curve, assuming the half-light radius reported in  \tabref{d6properties} and an ellipticity $\epsilon = 0$. This ellipticity is approximately matched to the mode of the marginalized posterior distribution for $\epsilon$ derived from our MCMC analysis; however, our constraint on the system's ellipticity ($\epsilon < 0.56$ at 95\% confidence) is relatively weak due to the small number of observed member stars.}
    \label{fig:mainmembership}
\end{figure*}

\subsection{Structural and Stellar Population Fit}
\label{sec:prop}

\begin{deluxetable}{l c c}
\tablewidth{0pt}
\tabletypesize{\footnotesize}
\tablecaption{\label{tab:d6properties} Properties of DELVE~6}
\tablehead{\colhead{Parameter} & \colhead{Value} & \colhead{Units}}
\startdata
IAU Name & DELVE J0212$-$6603 & ... \\
Constellation & Hydrus & ...\\
\ra & $33.070^{+0.003}_{-0.004}$ & deg \\
\dec & $-66.056^{+0.002}_{-0.002}$ & deg\\
$r_\text{h}$ & $0.43^{+0.18}_{-0.12}$ & arcmin  \\
$r_{1/2}$ & $10^{+4}_{-3}$ & pc  \\
\ellip & $< 0.56$ & ... \\
\PA & $14^{+40}_{-63}$ & deg \\
$M_V$\tablenotemark{a} & $-1.5^{+0.4}_{-0.6}$ & mag \\
\age & $> 9.8$ & Gyr \\
\feh & $< -1.17$ & dex \\
$(m-M)_0$ & $19.51^{+0.04}_{-0.12} \rm \ (stat.) \ \pm 0.1\tablenotemark{b}$ (sys.) & mag\\
$D_{\odot}$ & $80^{+2}_{-4} \rm \ (stat.) \  ^{+4}_{-4} \ (sys.)$ & kpc\\
$D_{\rm GC}$ & $78^{+2}_{-4} \rm \ (stat.) \  ^{+4}_{-4} \ (sys.)$ & kpc\\
$D_{\rm LMC}$ & $35^{+2}_{-3} \rm \ (stat.) \ ^{+3}_{-3} \ (sys.)$ & kpc\\
$D_{\rm SMC}$ & $20^{+2}_{-3} \rm \ (stat.) \  ^{+3}_{-3} \ (sys.)$ & kpc\\
$E(B-V)$\tablenotemark{c} & 0.036 & mag \\
\hline
$\mu_{\alpha} \cos \delta$ & $\ \ 0.93^{+0.39}_{-0.39}$ & mas yr$^{-1}$ \\
$\mu_{\delta}$ & $-1.28^{+0.38}_{-0.38}$ & mas yr$^{-1}$ \\
\enddata
\tablecomments{Uncertainties for each parameter were derived from the highest-density interval containing 68\% of the marginalized posterior distribution. For the ellipticity, metallicity, and age, the posterior distribution peaked at the boundary of the allowed parameter space. Therefore, we quote the upper, upper, and lower bound for these three parameters (respectively) at 95\% confidence.}
\tablenotetext{a}{Our estimate of $M_V$ was derived following the procedure from \citet{Martin:2008} and does not include uncertainty in the distance.}
\tablenotetext{b}{The statistical uncertainty on DELVE~6's distance is derived directly from our \ugali MCMC. We include a systematic uncertainty of $\pm0.1$ associated with isochrone modeling following \citet{2015ApJ...813..109D}. This systematic error is not included in the uncertainty on $r_{1/2}$.}
\tablenotetext{c}{This $E(B-V)$ value is the mean reddening of all sources within $r_{1/2}$, as determined via the maps of \cite{Schlegel:1998} with the recalibration from \citet{2011ApJ...737..103S}}
\end{deluxetable}

\par We fit DELVE 6's morphological and stellar population properties using the Ultra-faint Galaxy LIkelihood software toolkit (\ugali), which implements an unbinned Poisson maximum-likelihood approach based on the statistical formalism presented in Appendix C of \citet{2020ApJ...893...47D}. We modelled DELVE~6's structure with an elliptical \cite{1911MNRAS..71..460P} radial stellar density profile, and we fit a PARSEC isochrone \citep{Bressan:2012} to its observed $g,r$-band CMD. The eight free parameters for these models were the centroid coordinates ($\alpha_{\rm 2000}$ and $\delta_{\rm 2000}$), extension ($a_h$), ellipticity ($\epsilon$), position angle (P.A.) of the Plummer profile, and the age ($\tau$), metallicity ($\rm [Fe/H]$), and distance modulus ($(m-M)_0$) of the isochrone model. All eight of these parameters were constrained simultaneously by sampling their posterior probability distribution functions using the affine-invariant Markov Chain Monte Carlo ensemble sampler \citep{2010CAMCS...5...65G}  implemented in the Python package \code{emcee} \citep{Foreman-Mackey:2013}. This sampling was performed with 80 walkers each taking 35,000 steps, with the first 12,500 steps discarded as burn-in; these parameters were set to ensure dense sampling of the age--metallicity bimodality described later in the next subsection.
\par We report the best-fit values of each parameter and their associated uncertainties in \tabref{d6properties}. These results were derived from a fit assuming a nominal magnitude limit of $g_0, r_0 = 23.8$~mag, and with the size of the concentric annulus used by \ugali to construct the foreground/background model used for its joint fit of stellar color, magnitude, and spatial distributions set to $0.5\degree < r < 1.5\degree$. To assess whether these fit hyperparameters might affect our results given the complex, spatially-variable foreground/background stellar density associated with the MCs, we explored the sensitivity of our derived estimates of DELVE~6's properties to variations in the assumed magnitude limit and outer radius of the background annulus. Specifically, we re-ran the full \code{ugali} MCMC procedure for magnitude limits in the interval [23.6 mag, 24.0 mag] and background annulus radii in the interval [1.0$\degree$,2.0$\degree$].  In these tests, we found that all fits that converged resulted in parameter estimates consistent within uncertainties with those reported in \tabref{d6properties}; this was true for all parameters except the position angle, which is mostly unconstrained due to the negligible ellipticity of DELVE~6. However, we did need to apply a weak prior on the extension (sizes $0.001\degree < a_{h} < 0.1\degree$) in order to avoid convergence to non-physical results. Our fiducial results presented in \tabref{d6properties} were derived with this prior applied. 

\subsection{Properties of DELVE~6}
\par
\begin{figure*}
    \centering
    \includegraphics[width = .48\textwidth]{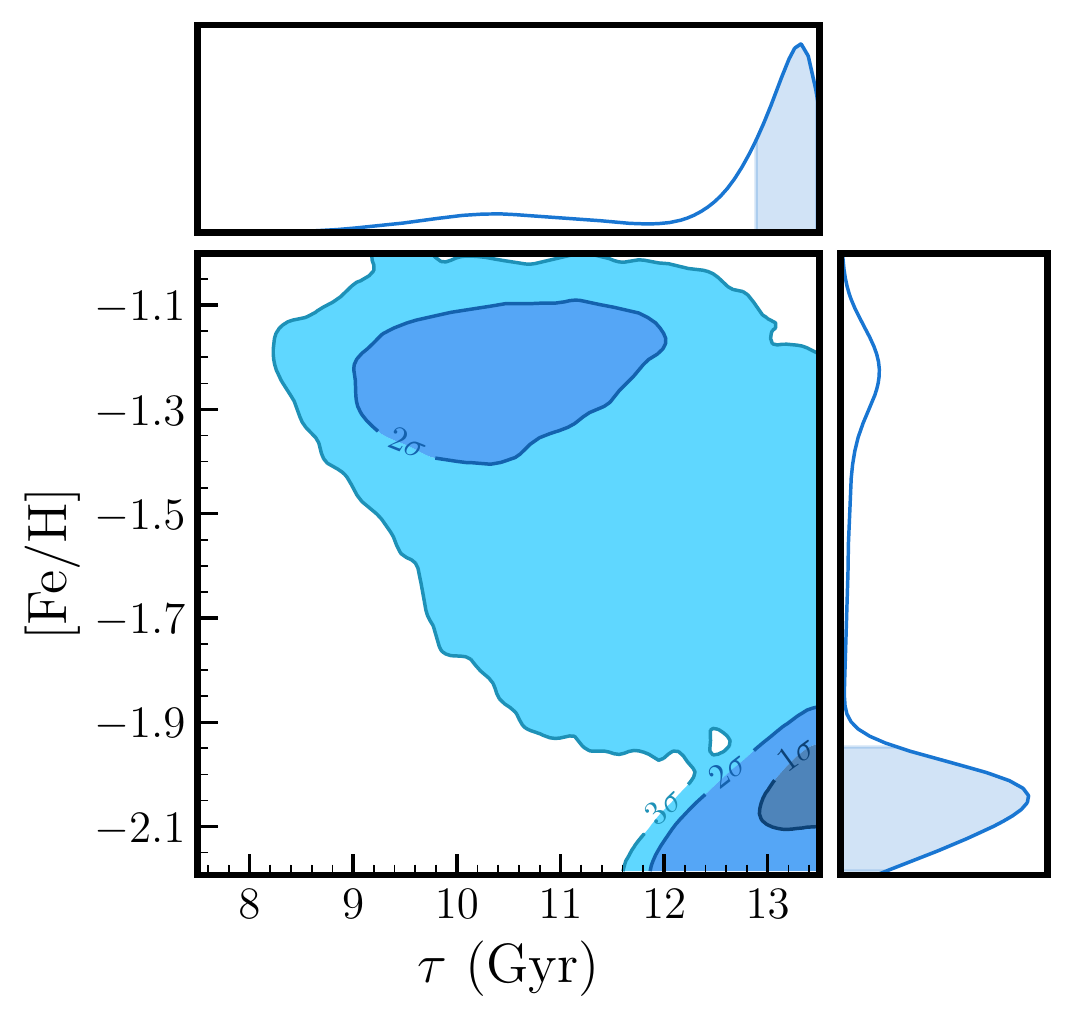}
     \raisebox{.015\height}{\includegraphics[width = .36\textwidth]{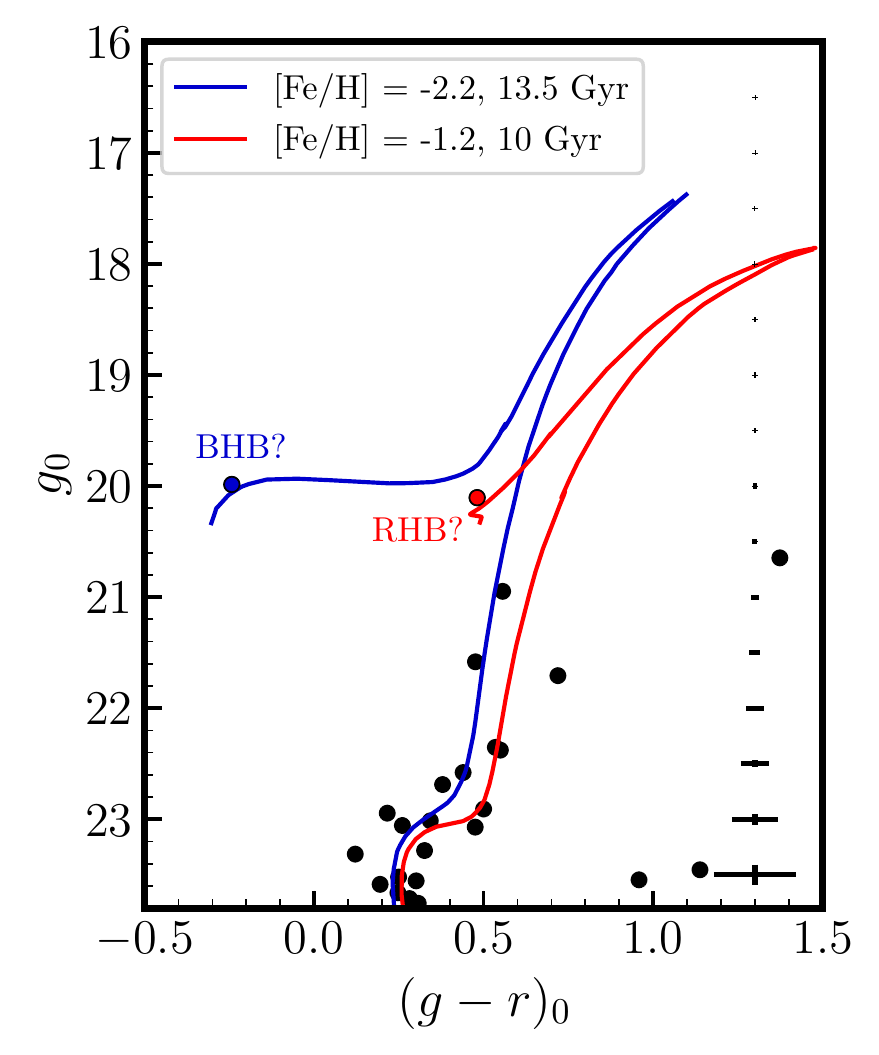}}
     \caption{(Left) 2D posterior probability distribution for the metallicity and age of DELVE~6. The blue shaded contours denote the $1\sigma, \ 2\sigma$, and $3\sigma$ 2D confidence regions in this plane. (Right) CMD for all stars within $2r_{1/2}$ of DELVE~6's centroid. Two isochrone models are shown. The blue model depicts an isochrone with ${\rm [Fe/H]} = -2.19$ and $\tau = 13.5$~Gyr, consistent with the posterior mode shown in the bottom-righthand corner of the left panel, whereas the red model depicts a younger, more metal-rich isochrone corresponding to a fit lying within the medium-blue $2\sigma$ contour near the top-center of the left panel. Here, both models are fixed to the distance modulus reported in \tabref{d6properties}, despite small differences in distance between the corresponding samples. We highlight that one BHB candidate lies along the older, more metal-poor (blue) model, and one (likely non-member) star lies near the RHB of the younger, more metal-rich (red) model.}
    \label{fig:amr}
\end{figure*}

 As depicted in \figref{mainmembership}--\ref{fig:amr}, we find that DELVE~6 is a compact ($r_{1/2} = 10^{+4}_{-3}$~pc), ultra-faint ($M_V = -1.5^{+0.4}_{-0.6}$ mag) stellar system with a round morphology ($\epsilon < 0.56$ at 95\%~confidence). These properties place DELVE~6 in a region of the $M_V$--$r_{1/2}$ plane that is dominated by the population of ultra-faint Milky Way halo star clusters discovered by recent surveys, which are generally fainter and more compact than their ultra-faint dwarf galaxy counterparts (see \figref{PopComp}). We do observe that DELVE~6 is among the more extended (candidate) ultra-faint star clusters discovered to date, though. We tentatively classify DELVE~6 as an ultra-faint star cluster on the basis of its small physical size, although a future spectroscopic measurement of its velocity and/or metallicity dispersion will provide a more definitive classification for the system.
\par In addition, we find that DELVE~6 is most consistent with an ancient stellar population, as evident from the clear main-sequence turnoff, extended subgiant branch, and (sparse) red giant branch seen in its observed CMD (\figref{mainmembership}--\ref{fig:amr}). The best-fit isochrone favored by our \ugali fit was consistent with the maximum age and the minimum metallicity of our isochrone grid ($\tau = 13.5$~Gyr and $\rm [Fe/H] = -2.19$~dex, respectively) although the marginalized posterior distribution for these two parameters was found to be bimodal, with a secondary peak at $\tau \sim 10$~Gyr and $\rm [Fe/H] \sim -1.2$~dex (see left panel of \figref{amr}). These modes appear to depend on whether the single blue horizontal branch (BHB) star candidate shown in the right panel of \figref{amr} is a true member, or alternately whether the star positioned near the red horizontal branch (RHB) of the second model is a true member. In our nominal best-fit parameter constraints, the BHB star is included as a high-confidence member, driving the fit toward the lower-metallicity mode (blue isochrone in \figref{amr}). Removing the BHB star from our photometric catalog and re-running the fit resulted in the secondary mode becoming the favored best-fit solution (red isochrone in the same figure). If we instead removed the candidate RHB star, the results are qualitatively similar to those shown in \figref{amr}.
\par Using the BHB--blue straggler separation technique introduced in \citet{2019MNRAS.490.3508L}, we found that the BHB star's $(g-r)_0$ and $(i-z)_0$ color is consistent with a classification as a \textit{bona fide} BHB star. Furthermore, the star's $\Gaia$ DR3 proper motion \citep{2022arXiv220800211G} is sufficiently small that we cannot clearly identify it as a foreground star, and its parallax is consistent with zero within errors; we therefore cannot rule out its membership in the distant DELVE~6 system. By contrast, the RHB candidate's proper motion of ($\mu_{\alpha}\cos \delta, \mu_{\delta}) = (-3.0 \pm 0.4, -12.2\pm 0.3$) implies a tangential velocity that is inconsistent with that of a typical star in a bound orbit at DELVE~6's distance; this suggests the star is not a member of DELVE~6. We have opted to present our \ugali fit results from a fit with both stars included in our catalog. This decision fully specifies the age/metallicity solution that we report for DELVE~6 due to the strong impact of the BHB candidate, but the upper/lower limits presented in the \tabref{d6properties} encompass both plausible solutions for these two parameters. Despite the uncertainty in the age and metallicity, we found that the structural properties presented in \tabref{d6properties} are consistent within uncertainties with the results derived from fits with either one of the two stars in question removed. 
\par Although we cannot conclusively determine which age/metallicity is most appropriate for DELVE~6 until deeper imaging and/or a spectroscopic metallicity measurement becomes available, both of these isochrone fits strongly suggest that the system is ancient ($\tau > 9.8$~Gyr). DELVE~6's old age, as well as its position in the outskirts of the LMC and SMC, raises interesting questions about its formation and evolution. We study the systemic proper motion of DELVE~6 below and then explore several possibilities for its origin in \secref{discuss}.

\subsection{Systemic Proper Motion of DELVE 6}
\label{sec:spm}
In addition to the aforementioned BHB and RHB candidates, we identified one additional nearby member candidate with a \Gaia proper motion measurement.
This star (\Gaia DR3 \code{source\_id}: 4698076296289956352) is the brightest RGB star consistent with our best-fit isochrone in \figref{mainmembership} and was identified as a high-probability member in our \ugali fit ($p_{\ugali} = 0.99$; see Appendix A, \tabref{PMs}). 
Notably, this RGB star’s proper motion is consistent with the BHB star at the $1.3 \sigma$ level.\footnote{Here, the confidence level
calculated appropriately for the case of two-dimensional Gaussians was
converted to the equivalent $\sigma$-distance for a one-dimensional
Gaussian; however, this calculation neglects the covariance between proper motion components.} The adequate agreement in proper motion supports the
interpretation that the BHB and RGB stars may both be members of DELVE 6, as the
joint probability of having these two stars by chance 
lying on a single isochrone while also sharing a statistically
consistent non-zero proper motion is small. 
It is thus reasonable to derive the proper motion of DELVE 6 from these two stars, which we find to be $(\mu_{\alpha} \cos \delta , \mu_\delta) = (0.93 \pm 0.39 , -1.28
\pm 0.38)$ mas yr$^{-1}$ through a simple two-parameter fit.

\par Although this measurement is tentative and should be interpreted cautiously, DELVE 6's kinematics are singularly important for unravelling its origins.  Therefore, we briefly explored what this measurement might tell us about the connection of DELVE 6 to possible host systems. To do so, we calculated its azimuthal and polar velocity components in Galactocentric spherical coordinates (hereafter $v_{\phi}$ and $v_{\theta}$) by sampling from the posterior distributions for the available 5D phase-space measurements and sampling the unknown radial velocity from a uniform distribution on the interval [$-500$ km $\rm s^{-1}$, 500 km $\rm s^{-1}$]. We then compared the values of these velocitity components to those expected from stars belonging to the distant MW halo and the MCs.
\par Using \code{galpy} \citep{2015ApJS..216...29B} to carry out the velocity transformation, we find $v_{\phi} = -50 \pm 150$ km $\rm s^{-1}$, $v_{\theta} = -430 \pm 150 $ km $\rm s^{-1}$, where the best-fit values and the upper/lower uncertainties correspond to the median and 16th/84th percentiles across 100,000 random 6D samples. This value of $v_{\phi}$ is uninformative, as it is consistent with the expected velocity distributions of all three hosts. However, we do observe that DELVE 6's estimated polar velocity $v_{\theta}$ is approximately consistent with that of the LMC  ($v_{\theta} \sim -305$  km $\rm s^{-1}$) and SMC ($v_{\theta} \sim -260$  km $\rm s^{-1}$). The polar velocity distribution of MW halo tracers at 80 kpc is expected to be a Gaussian centered near ${v_{\theta}} = 0\rm ~km~^{-1}$ with dispersion $\sigma_{v_{\theta}} \lesssim 100$~km~s$^{-1}$ (see Fig. 4 of \citealt{2019AJ....157..104B}), and thus DELVE 6 appears to be kinematically distinct from typical stars in the outer halo in this velocity component.
We therefore conclude that DELVE 6's proper motion, taken at face value, would argue in favor of a Magellanic association. This conclusion only weakly depends on the unknown line-of-sight velocity but is limited by the large error on DELVE 6's proper motion. Reduced proper motion errors (e.g., from future \Gaia data releases) will provide more definitive kinematic evidence for an association with one host or another.

\section{Discussion}
\label{sec:discuss}
At its projected sky position (\figref{cutout}) and Galactocentric distance ($D_{\rm GC} = 78$~kpc), DELVE~6 is located $20$ kpc from the center of the SMC and $35$~kpc from the center of the LMC.\footnote{We assume Galactocentric distances for the LMC, SMC, and Galactic center from \citet{2013Natur.495...76P,2020ApJ...904...13G} and \citet{2019A&A...625L..10G}, respectively. We neglect the uncertainties on these distances and on each object's centroid coordinates when calculating $D_{\rm LMC}$, $D_{\rm SMC}, D_{\rm GC}$ because they are subdominant relative to the uncertainty on DELVE~6's heliocentric distance. } Despite this relative proximity, DELVE~6 is located beyond the tidal radius ($r_t$) of the LMC due to the MW ($r_t \sim 22$~kpc; \citealt{2014ApJ...781..121V}) and the tidal radius of the SMC due to the LMC ($r_t \sim 5$~kpc; \citealt{2020MNRAS.498.1034M}), suggesting that these two possible hosts have a weak gravitational influence on DELVE~6. It is therefore not immediately evident based on its position alone whether this system is associated with either of the Clouds. On the basis of its separation from the LMC and SMC, as well as the possible ages and metallicities discussed above, we speculate that DELVE 6 is most likely described by one of three scenarios: (1) a distant ultra-faint Milky Way halo star cluster coincidentally located near the MCs; (2) a LMC cluster residing in its host's outer halo in a weakly-bound or unbound orbit, or (3) a cluster formed within the SMC that has been stripped from its host and now orbits in the MW+LMC potential. 

\begin{figure}
    \centering
    \includegraphics[width = .47\textwidth]{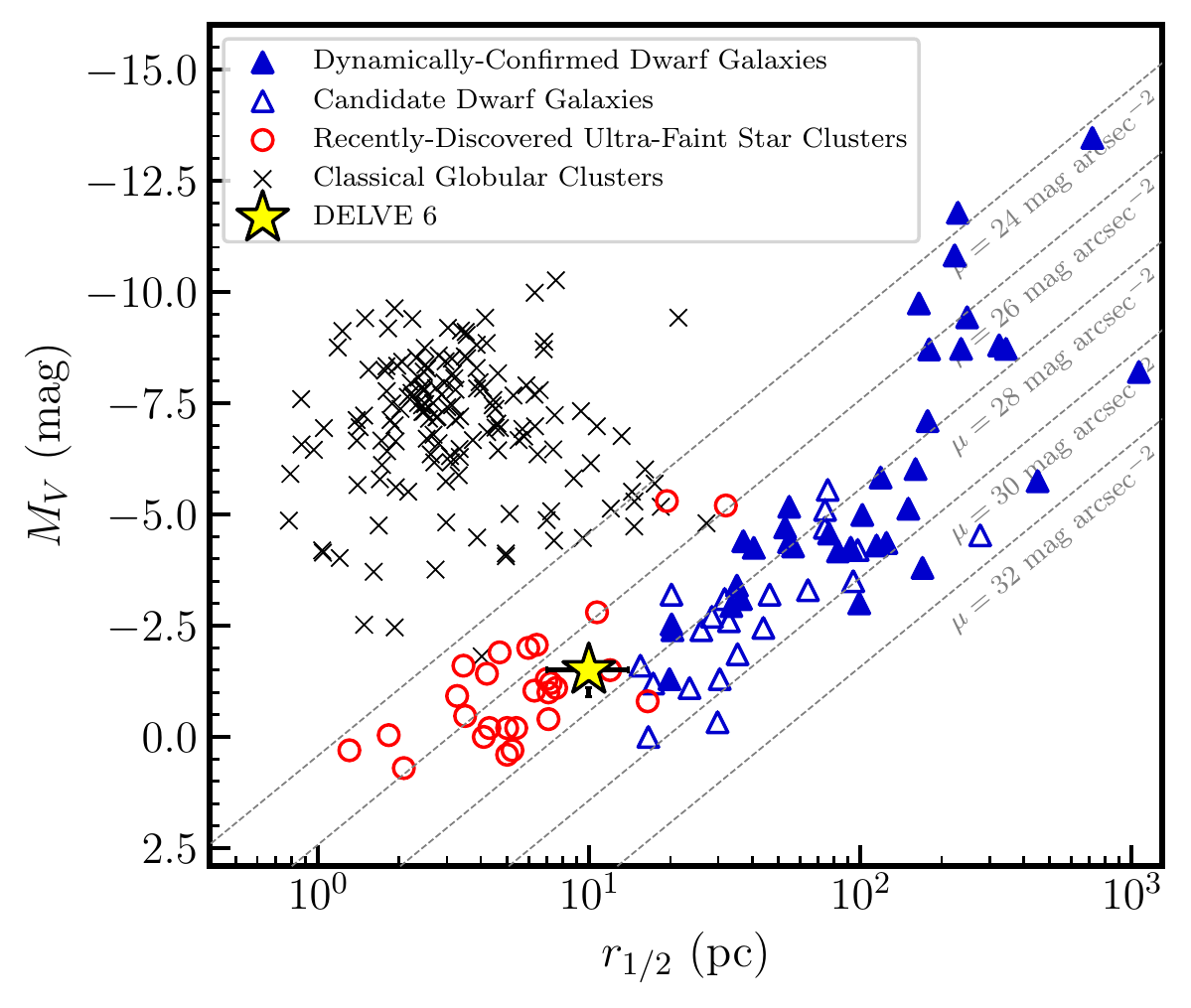}
    \caption{Absolute magnitude ($M_V$) vs.\ azimuthally-averaged half-light radius ($r_{1/2}$) for a large sample of classical globular clusters, candidate and confirmed Milky Way satellite galaxies, and recently-discovered ultra-faint halo star clusters. The location of DELVE 6 in this plane is indicated by a yellow star. A complete reference list for this figure is available in Appendix \ref{sec:refs}.}
    \label{fig:PopComp}
\end{figure}

\par The first of these scenarios is supported by the similarity between DELVE 6's age and metallicity to the properties of the MW's ``classical'' globular clusters and the growing population of ultra-faint MW halo star clusters. This scenario also mitigates the need for an explanation of DELVE 6's apparently large separation from the LMC and SMC. Roughly a dozen MW star clusters are known at $D_{\odot} > 70$~kpc, suggesting that an LMC/SMC origin is not necessary to explain DELVE~6's large Galactocentric distance.\footnote{This number includes ultra-faint systems, but excludes clusters in the MW's dwarf satellites (e.g. Fornax and Eridanus II).} This all being said, our exploration of DELVE 6's tentative proper motion (\secref{spm}) suggests that its overall kinematics may be more consistent with the LMC/SMC system and inconsistent with MW halo tracers at its distance. 
\par On that note, the second of our proposed scenarios -- namely that the system was formed in the LMC and remains in a weakly-bound or unbound orbit around its host -- may also explain the ancient age of DELVE~6 given the $\sim 15$ known LMC globular clusters with ages $> 9$~Gyr \citep[e.g.,][]{2004MNRAS.352..153M}, but does not directly explain the system's present-day Galactocentric distance and 3D separation. Indeed, if confirmed to be an LMC satellite, DELVE~6 would reside at a larger separation from the LMC than all but two star clusters believed to be associated with the MCs.\footnote{The two candidate LMC/SMC clusters at larger separations are DES~1 \citep{2016MNRAS.458..603L} and Eridanus~III \citep{Bechtol:2015}; both are included in \figref{cutout}. \citet{2018ApJ...852...68C} argue in favor of a Magellanic association for both, although their analysis did not rely on any kinematic evidence.}
Nevertheless, there is evidence that the LMC dark matter halo extends $>50$ kpc in radius \citep[e.g.,][]{2022arXiv221104495K}, making it plausible that DELVE~6 lies in the outskirts of the LMC halo. Furthermore, we note that at least two ultra-faint dwarf galaxies that are likely to be associated with the LMC lie at larger separations from their (original) host compared to DELVE~6: Horologium I and Phoenix II \citep[see][]{2022ApJ...940..136P}. Detailed modelling of the MW and LMC dark matter halo structures (including the LMC's dynamical friction wake) and these satellites' orbits within the associated potential suggests that the former of these two ultra-faint dwarfs is unbound from the LMC, while the latter is likely bound \citep{2021ApJ...919..109G}. By analogy, we conclude that DELVE~6 could plausibly be either weakly bound to the LMC or unbound given its current position. 
\par The last possibility is that DELVE~6 was initially formed within the SMC, but was stripped from its host due to interactions between the LMC and the SMC. The strongest (and arguably only) observational evidence for this conclusion is DELVE~6's large present-day Galactocentric distance and its on-sky location. These properties place the system in a 3D position where a relatively high density of SMC satellites/debris is expected based on numerical simulations \citep{2016MNRAS.461.2212J}. In further support of this possibility, we highlight that a similar stripping scenario has also been proposed for the recently-discovered ultra-faint star cluster YMCA-1 on the basis of its proper motion \citep{2022MNRAS.515.4005P}; like DELVE 6, this system lies well beyond the SMC's tidal radius.

\par Disfavoring this stripping scenario is the fact that DELVE~6's age and metallicity from our nominal best-fit solution are inconsistent with the known SMC globular cluster population, which includes only a single comparably old system with a robustly-measured age (NGC~121, at $\tau \sim 11$~Gyr; \citealt{2008AJ....135.1106G}). Indeed, our measurement suggests that DELVE~6 would be the second or third oldest star cluster associated with the SMC, depending on the status of the aforementioned YMCA-1 system, which has a somewhat uncertain age ($\tau \sim 9.6 - 11.7$ Gyr; \citealt{2022MNRAS.515.4005P,2022ApJ...929L..21G}) and only a tentative association with the SMC. Nonetheless, such a stripping scenario has been proposed as one explanation for the apparent dearth of old-aged star clusters associated with the SMC. By tracing the orbits of star clusters throughout a period of dynamical interaction between the LMC and SMC, \citet{2013MNRAS.435L..63C} found that for large eccentricities ($e > 0.5$), $\sim 15\%$ of SMC star clusters are captured by the LMC and an additional $\sim 20-50\%$ of clusters are ejected into the intergalactic medium.
We believe that either of these two capture scenarios is more likely than the possibility that DELVE~6 inhabits a weakly-bound orbit in the outer halo of the SMC (analogous to the case above for the LMC) given that the system lies at $\sim 4 r_{\rm t,SMC}$ compared to $\sim 1.6 r_{\rm t, LMC}$.  

\par Spectroscopic follow-up will be critical for distinguishing between these scenarios. Specifically, a radial velocity measurement would enable the possibility of rewinding DELVE~6's orbit in the combined MW+LMC+SMC potential, elucidating its origins. If an LMC or SMC origin can be robustly established on the basis of its orbit, DELVE~6 may provide a significant constraint on its host's age-metallicity relation at a very large radius. Deeper imaging reaching the main-sequence turnoff feature of DELVE~6's color--magnitude diagram will be necessary to realize this possibility and robustly confirm this system's ancient age, and a spectroscopic metallicity measurement would aid in breaking the age-metallicity degeneracy inherent to isochrone fitting.

\par Lastly, we highlight that the discovery of DELVE~6 emphasizes that the observational census of ultra-faint systems in the Magellanic environment is incomplete. Considering the continual discovery of similar ultra-faint star cluster systems near the MCs in recent years, we speculate that a more extensive population of old, metal-poor ultra-faint star clusters may exist around the LMC, SMC, and perhaps even other low-mass galaxies in the Local Group, waiting to be uncovered by current and future surveys.

\section{Acknowledgments}
Codes and data products associated with this work are available online at \url{https://github.com/wcerny/DELVE6\_Paper}.
\par This project is partially supported by the NASA Fermi
Guest Investigator Program Cycle 9 No. 91201. 
This work is partially supported by Fermilab LDRD project
L2019-011.
W.C. gratefully acknowledges support from
a Gruber Science Fellowship at Yale University. 
CEMV is supported by the international Gemini Observatory, a program of NSF’s NOIRLab, which is managed by the Association of Universities for Research in Astronomy (AURA) under a cooperative agreement with the National Science Foundation, on behalf of the Gemini partnership of Argentina, Brazil, Canada, Chile, the Republic of Korea, and the United States of America. JAC-B acknowledges support from FONDECYT Regular N 1220083.


This project used data obtained with the Dark Energy Camera, which was constructed by the Dark Energy Survey (DES) collaboration.
Funding for the DES Projects has been provided by 
the DOE and NSF (USA),   
MISE (Spain),   
STFC (UK), 
HEFCE (UK), 
NCSA (UIUC), 
KICP (U. Chicago), 
CCAPP (Ohio State), 
MIFPA (Texas A\&M University),  
CNPQ, 
FAPERJ, 
FINEP (Brazil), 
MINECO (Spain), 
DFG (Germany), 
and the collaborating institutions in the Dark Energy Survey, which are
Argonne Lab, 
UC Santa Cruz, 
University of Cambridge, 
CIEMAT-Madrid, 
University of Chicago, 
University College London, 
DES-Brazil Consortium, 
University of Edinburgh, 
ETH Z{\"u}rich, 
Fermilab, 
University of Illinois, 
ICE (IEEC-CSIC), 
IFAE Barcelona, 
Lawrence Berkeley Lab, 
LMU M{\"u}nchen, and the associated Excellence Cluster Universe, 
University of Michigan, 
NSF's National Optical-Infrared Astronomy Research Laboratory, 
University of Nottingham, 
Ohio State University, 
OzDES Membership Consortium
University of Pennsylvania, 
University of Portsmouth, 
SLAC National Lab, 
Stanford University, 
University of Sussex, 
and Texas A\&M University.

Based on observations at Cerro Tololo Inter-American Observatory, NSF's National Optical-Infrared Astronomy Research Laboratory (2019A-0305; PI: Drlica-Wagner), which is operated by the Association of Universities for Research in Astronomy (AURA) under a cooperative agreement with the National Science Foundation.

The Legacy Surveys consist of three individual and complementary projects: the Dark Energy Camera Legacy Survey (DECaLS; Proposal ID 2014B-0404; PIs: David Schlegel and Arjun Dey), the Beijing-Arizona Sky Survey (BASS; NOAO Prop. ID 2015A-0801; PIs: Zhou Xu and Xiaohui Fan), and the Mayall z-band Legacy Survey (MzLS; Prop. ID 2016A-0453; PI: Arjun Dey). DECaLS, BASS and MzLS together include data obtained, respectively, at the Blanco telescope, Cerro Tololo Inter-American Observatory, NSF’s NOIRLab; the Bok telescope, Steward Observatory, University of Arizona; and the Mayall telescope, Kitt Peak National Observatory, NOIRLab. Pipeline processing and analyses of the data were supported by NOIRLab and the Lawrence Berkeley National Laboratory (LBNL). The Legacy Surveys project is honored to be permitted to conduct astronomical research on Iolkam Du’ag (Kitt Peak), a mountain with particular significance to the Tohono O’odham Nation.

NOIRLab is operated by the Association of Universities for Research in Astronomy (AURA) under a cooperative agreement with the National Science Foundation. LBNL is managed by the Regents of the University of California under contract to the U.S. Department of Energy.

BASS is a key project of the Telescope Access Program (TAP), which has been funded by the National Astronomical Observatories of China, the Chinese Academy of Sciences (the Strategic Priority Research Program “The Emergence of Cosmological Structures” Grant \# XDB09000000), and the Special Fund for Astronomy from the Ministry of Finance. The BASS is also supported by the External Cooperation Program of Chinese Academy of Sciences (Grant \# 114A11KYSB20160057), and Chinese National Natural Science Foundation (Grant \# 12120101003, \# 11433005).

The Legacy Survey team makes use of data products from the Near-Earth Object Wide-field Infrared Survey Explorer (NEOWISE), which is a project of the Jet Propulsion Laboratory/California Institute of Technology. NEOWISE is funded by the National Aeronautics and Space Administration.

The Legacy Surveys imaging of the DESI footprint is supported by the Director, Office of Science, Office of High Energy Physics of the U.S. Department of Energy under Contract No. DE-AC02-05CH1123, by the National Energy Research Scientific Computing Center, a DOE Office of Science User Facility under the same contract; and by the U.S. National Science Foundation, Division of Astronomical Sciences under Contract No. AST-0950945 to NOAO.

This work has made use of data from the European Space Agency (ESA) mission {\it Gaia} (\url{https://www.cosmos.esa.int/gaia}), processed by the {\it Gaia} Data Processing and Analysis Consortium (DPAC, \url{https://www.cosmos.esa.int/web/gaia/dpac/consortium}).
Funding for the DPAC has been provided by national institutions, in particular the institutions participating in the {\it Gaia} Multilateral Agreement.

\par This work made use of Astropy:\footnote{http://www.astropy.org} a community-developed core Python package and an ecosystem of tools and resources for astronomy.

This manuscript has been authored by Fermi Research Alliance, LLC, under contract No.\ DE-AC02-07CH11359 with the US Department of Energy, Office of Science, Office of High Energy Physics. The United States Government retains and the publisher, by accepting the article for publication, acknowledges that the United States Government retains a non-exclusive, paid-up, irrevocable, worldwide license to publish or reproduce the published form of this manuscript, or allow others to do so, for United States Government purposes.

\facility{Blanco, \Gaia.}
\software{\code{numpy} \citep{2011CSE....13b..22V,2020Natur.585..357H}, \code{scipy} \citep{2020NatMe..17..261V}, \emcee \citep{Foreman-Mackey:2013}, \healpix \citep{2005ApJ...622..759G},\footnote{\url{http://healpix.sourceforge.net}} \code{healpy} \citep{2019JOSS....4.1298Z} , \ugali \citep{Bechtol:2015} \footnote{\url{https://github.com/DarkEnergySurvey/ugali}}, \code{ChainConsumer} \citep{2019ascl.soft10017H}, \code{simple} \citep{Bechtol:2015,2015ApJ...813..109D}, \code{astropy} \citep{2013A&A...558A..33A,2018AJ....156..123A,2022ApJ...935..167A}}

\bibliography{main}

\begin{thebibliography}{}
\expandafter\ifx\csname natexlab\endcsname\relax\def\natexlab#1{#1}\fi
\providecommand{\url}[1]{\href{#1}{#1}}

\bibitem[{{Astropy Collaboration} {et~al.}(2013){Astropy Collaboration},
  {Robitaille}, {Tollerud}, {Greenfield}, {Droettboom}, {Bray}, {Aldcroft},
  {Davis}, {Ginsburg}, {Price-Whelan}, {Kerzendorf}, {Conley}, {Crighton},
  {Barbary}, {Muna}, {Ferguson}, {Grollier}, {Parikh}, {Nair}, {Unther},
  {Deil}, {Woillez}, {Conseil}, {Kramer}, {Turner}, {Singer}, {Fox}, {Weaver},
  {Zabalza}, {Edwards}, {Azalee Bostroem}, {Burke}, {Casey}, {Crawford},
  {Dencheva}, {Ely}, {Jenness}, {Labrie}, {Lim}, {Pierfederici}, {Pontzen},
  {Ptak}, {Refsdal}, {Servillat}, \& {Streicher}}]{2013A&A...558A..33A}
{Astropy Collaboration}, {Robitaille}, T.~P., {Tollerud}, E.~J., {et~al.} 2013,
  \aap, 558, A33

\bibitem[{{Astropy Collaboration} {et~al.}(2018){Astropy Collaboration},
  {Price-Whelan}, {Sip{\H{o}}cz}, {G{\"u}nther}, {Lim}, {Crawford}, {Conseil},
  {Shupe}, {Craig}, {Dencheva}, {Ginsburg}, {VanderPlas}, {Bradley},
  {P{\'e}rez-Su{\'a}rez}, {de Val-Borro}, {Aldcroft}, {Cruz}, {Robitaille},
  {Tollerud}, {Ardelean}, {Babej}, {Bach}, {Bachetti}, {Bakanov}, {Bamford},
  {Barentsen}, {Barmby}, {Baumbach}, {Berry}, {Biscani}, {Boquien}, {Bostroem},
  {Bouma}, {Brammer}, {Bray}, {Breytenbach}, {Buddelmeijer}, {Burke},
  {Calderone}, {Cano Rodr{\'\i}guez}, {Cara}, {Cardoso}, {Cheedella}, {Copin},
  {Corrales}, {Crichton}, {D'Avella}, {Deil}, {Depagne}, {Dietrich}, {Donath},
  {Droettboom}, {Earl}, {Erben}, {Fabbro}, {Ferreira}, {Finethy}, {Fox},
  {Garrison}, {Gibbons}, {Goldstein}, {Gommers}, {Greco}, {Greenfield},
  {Groener}, {Grollier}, {Hagen}, {Hirst}, {Homeier}, {Horton}, {Hosseinzadeh},
  {Hu}, {Hunkeler}, {Ivezi{\'c}}, {Jain}, {Jenness}, {Kanarek}, {Kendrew},
  {Kern}, {Kerzendorf}, {Khvalko}, {King}, {Kirkby}, {Kulkarni}, {Kumar},
  {Lee}, {Lenz}, {Littlefair}, {Ma}, {Macleod}, {Mastropietro}, {McCully},
  {Montagnac}, {Morris}, {Mueller}, {Mumford}, {Muna}, {Murphy}, {Nelson},
  {Nguyen}, {Ninan}, {N{\"o}the}, {Ogaz}, {Oh}, {Parejko}, {Parley}, {Pascual},
  {Patil}, {Patil}, {Plunkett}, {Prochaska}, {Rastogi}, {Reddy Janga},
  {Sabater}, {Sakurikar}, {Seifert}, {Sherbert}, {Sherwood-Taylor}, {Shih},
  {Sick}, {Silbiger}, {Singanamalla}, {Singer}, {Sladen}, {Sooley},
  {Sornarajah}, {Streicher}, {Teuben}, {Thomas}, {Tremblay}, {Turner},
  {Terr{\'o}n}, {van Kerkwijk}, {de la Vega}, {Watkins}, {Weaver}, {Whitmore},
  {Woillez}, {Zabalza}, \& {Astropy Contributors}}]{2018AJ....156..123A}
{Astropy Collaboration}, {Price-Whelan}, A.~M., {Sip{\H{o}}cz}, B.~M., {et~al.}
  2018, \aj, 156, 123

\bibitem[{{Astropy Collaboration} {et~al.}(2022){Astropy Collaboration},
  {Price-Whelan}, {Lim}, {Earl}, {Starkman}, {Bradley}, {Shupe}, {Patil},
  {Corrales}, {Brasseur}, {N{\"o}the}, {Donath}, {Tollerud}, {Morris},
  {Ginsburg}, {Vaher}, {Weaver}, {Tocknell}, {Jamieson}, {van Kerkwijk},
  {Robitaille}, {Merry}, {Bachetti}, {G{\"u}nther}, {Aldcroft},
  {Alvarado-Montes}, {Archibald}, {B{\'o}di}, {Bapat}, {Barentsen},
  {Baz{\'a}n}, {Biswas}, {Boquien}, {Burke}, {Cara}, {Cara}, {Conroy},
  {Conseil}, {Craig}, {Cross}, {Cruz}, {D'Eugenio}, {Dencheva}, {Devillepoix},
  {Dietrich}, {Eigenbrot}, {Erben}, {Ferreira}, {Foreman-Mackey}, {Fox},
  {Freij}, {Garg}, {Geda}, {Glattly}, {Gondhalekar}, {Gordon}, {Grant},
  {Greenfield}, {Groener}, {Guest}, {Gurovich}, {Handberg}, {Hart},
  {Hatfield-Dodds}, {Homeier}, {Hosseinzadeh}, {Jenness}, {Jones}, {Joseph},
  {Kalmbach}, {Karamehmetoglu}, {Ka{\l}uszy{\'n}ski}, {Kelley}, {Kern},
  {Kerzendorf}, {Koch}, {Kulumani}, {Lee}, {Ly}, {Ma}, {MacBride}, {Maljaars},
  {Muna}, {Murphy}, {Norman}, {O'Steen}, {Oman}, {Pacifici}, {Pascual},
  {Pascual-Granado}, {Patil}, {Perren}, {Pickering}, {Rastogi}, {Roulston},
  {Ryan}, {Rykoff}, {Sabater}, {Sakurikar}, {Salgado}, {Sanghi}, {Saunders},
  {Savchenko}, {Schwardt}, {Seifert-Eckert}, {Shih}, {Jain}, {Shukla}, {Sick},
  {Simpson}, {Singanamalla}, {Singer}, {Singhal}, {Sinha}, {Sip{\H{o}}cz},
  {Spitler}, {Stansby}, {Streicher}, {{\v{S}}umak}, {Swinbank}, {Taranu},
  {Tewary}, {Tremblay}, {Val-Borro}, {Van Kooten}, {Vasovi{\'c}}, {Verma}, {de
  Miranda Cardoso}, {Williams}, {Wilson}, {Winkel}, {Wood-Vasey}, {Xue},
  {Yoachim}, {Zhang}, {Zonca}, \& {Astropy Project
  Contributors}}]{2022ApJ...935..167A}
{Astropy Collaboration}, {Price-Whelan}, A.~M., {Lim}, P.~L., {et~al.} 2022,
  \apj, 935, 167

\bibitem[{{Balbinot} {et~al.}(2013){Balbinot}, {Santiago}, {da Costa}, {Maia},
  {Majewski}, {Nidever}, {Rocha-Pinto}, {Thomas}, {Wechsler}, \&
  {Yanny}}]{2013ApJ...767..101B}
{Balbinot}, E., {Santiago}, B.~X., {da Costa}, L., {et~al.} 2013, \apj, 767,
  101

\bibitem[{{Battaglia} {et~al.}(2022){Battaglia}, {Taibi}, {Thomas}, \&
  {Fritz}}]{2022A&A...657A..54B}
{Battaglia}, G., {Taibi}, S., {Thomas}, G.~F., \& {Fritz}, T.~K. 2022, \aap,
  657, A54

\bibitem[{{Bechtol}(2017)}]{2017AAS...22941606B}
{Bechtol}, K. 2017, in American Astronomical Society Meeting Abstracts, Vol.
  229, American Astronomical Society Meeting Abstracts \#229, 416.06

\bibitem[{{Bechtol} {et~al.}(2015){Bechtol}, {Drlica-Wagner}, {Balbinot},
  {Pieres}, {Simon}, {Yanny}, {Santiago}, {Wechsler}, {Frieman}, {Walker},
  {Williams}, {Rozo}, {Rykoff}, {Queiroz}, {Luque}, {Benoit-L{\'e}vy},
  {Tucker}, {Sevilla}, {Gruendl}, {da Costa}, {Fausti Neto}, {Maia}, {Abbott},
  {Allam}, {Armstrong}, {Bauer}, {Bernstein}, {Bernstein}, {Bertin}, {Brooks},
  {Buckley-Geer}, {Burke}, {Carnero Rosell}, {Castander}, {Covarrubias},
  {D'Andrea}, {DePoy}, {Desai}, {Diehl}, {Eifler}, {Estrada}, {Evrard},
  {Fernandez}, {Finley}, {Flaugher}, {Gaztanaga}, {Gerdes}, {Girardi},
  {Gladders}, {Gruen}, {Gutierrez}, {Hao}, {Honscheid}, {Jain}, {James},
  {Kent}, {Kron}, {Kuehn}, {Kuropatkin}, {Lahav}, {Li}, {Lin}, {Makler},
  {March}, {Marshall}, {Martini}, {Merritt}, {Miller}, {Miquel}, {Mohr},
  {Neilsen}, {Nichol}, {Nord}, {Ogando}, {Peoples}, {Petravick}, {Plazas},
  {Romer}, {Roodman}, {Sako}, {Sanchez}, {Scarpine}, {Schubnell}, {Smith},
  {Soares-Santos}, {Sobreira}, {Suchyta}, {Swanson}, {Tarle}, {Thaler},
  {Thomas}, {Wester}, {Zuntz}, \& {DES Collaboration}}]{Bechtol:2015}
{Bechtol}, K., {Drlica-Wagner}, A., {Balbinot}, E., {et~al.} 2015, \apj, 807,
  50

\bibitem[{{Bica} {et~al.}(2008){Bica}, {Bonatto}, {Dutra}, \&
  {Santos}}]{2008MNRAS.389..678B}
{Bica}, E., {Bonatto}, C., {Dutra}, C.~M., \& {Santos}, J.~F.~C. 2008, \mnras,
  389, 678

\bibitem[{{Bica} {et~al.}(2020){Bica}, {Westera}, {Kerber}, {Dias}, {Maia},
  {Santos}, {Barbuy}, \& {Oliveira}}]{2020AJ....159...82B}
{Bica}, E., {Westera}, P., {Kerber}, L. d.~O., {et~al.} 2020, \aj, 159, 82

\bibitem[{{Bird} {et~al.}(2019){Bird}, {Xue}, {Liu}, {Shen}, {Flynn}, \&
  {Yang}}]{2019AJ....157..104B}
{Bird}, S.~A., {Xue}, X.-X., {Liu}, C., {et~al.} 2019, \aj, 157, 104

\bibitem[{{Bitsakis} {et~al.}(2018){Bitsakis}, {Gonz{\'a}lez-L{\'o}pezlira},
  {Bonfini}, {Bruzual}, {Maravelias}, {Zaritsky}, {Charlot}, \&
  {Ram{\'\i}rez-Siordia}}]{2018ApJ...853..104B}
{Bitsakis}, T., {Gonz{\'a}lez-L{\'o}pezlira}, R.~A., {Bonfini}, P., {et~al.}
  2018, \apj, 853, 104

\bibitem[{{Bovy}(2015)}]{2015ApJS..216...29B}
{Bovy}, J. 2015, \apjs, 216, 29

\bibitem[{{Bressan} {et~al.}(2012){Bressan}, {Marigo}, {Girardi}, {Salasnich},
  {Dal Cero}, {Rubele}, \& {Nanni}}]{Bressan:2012}
{Bressan}, A., {Marigo}, P., {Girardi}, L., {et~al.} 2012, \mnras, 427, 127

\bibitem[{{Cantu} {et~al.}(2021){Cantu}, {Pace}, {Marshall}, {Strigari},
  {Crnojevic}, {Simon}, {Drlica-Wagner}, {Bechtol},
  {Mart{\'\i}nez-V{\'a}zquez}, {Santiago}, {Amara}, {Stringer}, {Diehl},
  {Aguena}, {Allam}, {Avila}, {Brooks}, {Carnero Rosell}, {Carrasco Kind},
  {Carretero}, {Costanzi}, {Da Costa}, {De Vicente}, {Desai}, {Doel}, {Eifler},
  {Everett}, {Frieman}, {Garc{\'\i}a-Bellido}, {Gaztanaga}, {Gruen}, {Gruendl},
  {Gschwend}, {Gutierrez}, {Hinton}, {Hollowood}, {Honscheid}, {James},
  {Kuehn}, {Maia}, {Menanteau}, {Miquel}, {Palmese}, {Paz-Chinch{\'o}n},
  {Plazas}, {Sanchez}, {Scarpine}, {Schubnell}, {Serrano}, {Sevilla-Noarbe},
  {Smith}, {Soares-Santos}, {Suchyta}, {Swanson}, {Tarle}, {Walker},
  {Wilkinson}, \& {DES Collaboration}}]{2021ApJ...916...81C}
{Cantu}, S.~A., {Pace}, A.~B., {Marshall}, J., {et~al.} 2021, \apj, 916, 81

\bibitem[{{Carlin} {et~al.}(2017){Carlin}, {Sand}, {Mu{\~n}oz}, {Spekkens},
  {Willman}, {Crnojevi{\'c}}, {Forbes}, {Hargis}, {Kirby}, {Peter},
  {Romanowsky}, \& {Strader}}]{2017AJ....154..267C}
{Carlin}, J.~L., {Sand}, D.~J., {Mu{\~n}oz}, R.~R., {et~al.} 2017, \aj, 154,
  267

\bibitem[{{Carpintero} {et~al.}(2013){Carpintero}, {Gomez}, \&
  {Piatti}}]{2013MNRAS.435L..63C}
{Carpintero}, D.~D., {Gomez}, F.~A., \& {Piatti}, A.~E. 2013, \mnras, 435, L63

\bibitem[{{Cerny} {et~al.}(2021{\natexlab{a}}){Cerny}, {Pace}, {Drlica-Wagner},
  {Ferguson}, {Mau}, {Adam{\'o}w}, {Carlin}, {Choi}, {Erkal}, {Johnson}, {Li},
  {Mart{\'\i}nez-V{\'a}zquez}, {Mutlu-Pakdil}, {Nidever}, {Olsen}, {Pieres},
  {Tollerud}, {Simon}, {Vivas}, {James}, {Kuropatkin}, {Majewski},
  {Mart{\'\i}nez-Delgado}, {Massana}, {Miller}, {Neilsen}, {No{\"e}l}, {Riley},
  {Sand}, {Santana-Silva}, {Stringfellow}, {Tucker}, \& {Delve
  Collaboration}}]{2021ApJ...910...18C}
{Cerny}, W., {Pace}, A.~B., {Drlica-Wagner}, A., {et~al.} 2021{\natexlab{a}},
  \apj, 910, 18

\bibitem[{{Cerny} {et~al.}(2021{\natexlab{b}}){Cerny}, {Pace}, {Drlica-Wagner},
  {Koposov}, {Vivas}, {Mau}, {Riley}, {Bom}, {Carlin}, {Choi}, {Erkal},
  {Ferguson}, {James}, {Li}, {Mart{\'\i}nez-Delgado},
  {Mart{\'\i}nez-V{\'a}zquez}, {Munoz}, {Mutlu-Pakdil}, {Olsen}, {Pieres},
  {Sakowska}, {Sand}, {Simon}, {Smercina}, {Stringfellow}, {Tollerud},
  {Adam{\'o}w}, {Hernandez-Lang}, {Kuropatkin}, {Santana-Silva}, {Tucker},
  {Zenteno}, \& {Delve Collaboration}}]{2021ApJ...920L..44C}
---. 2021{\natexlab{b}}, \apjl, 920, L44

\bibitem[{{Cerny} {et~al.}(2022){Cerny}, {Mart{\'\i}nez-V{\'a}zquez},
  {Drlica-Wagner}, {Pace}, {Mutlu-Pakdil}, {Li}, {Riley}, {Crnojevi{\'c}},
  {Bom}, {Carballo-Bello}, {Carlin}, {Chiti}, {Choi}, {Collins},
  {Darragh-Ford}, {Ferguson}, {Geha}, {Mart{\'\i}nez-Delgado}, {Massana},
  {Mau}, {Medina}, {Mu{\~n}oz}, {Nadler}, {Olsen}, {Pieres}, {Sakowska},
  {Simon}, {Stringfellow}, {Vivas}, {Walker}, \&
  {Wechsler}}]{2022arXiv220912422C}
{Cerny}, W., {Mart{\'\i}nez-V{\'a}zquez}, C.~E., {Drlica-Wagner}, A., {et~al.}
  2022, arXiv e-prints, arXiv:2209.12422

\bibitem[{{Cerny} {et~al.}(2023){Cerny}, {Simon}, {Li}, {Drlica-Wagner},
  {Pace}, {Mart{\'\i}nez-V{\'a}zquez}, {Riley}, {Mutlu-Pakdil}, {Mau},
  {Ferguson}, {Erkal}, {Munoz}, {Bom}, {Carlin}, {Carollo}, {Choi}, {Ji},
  {Manwadkar}, {Mart{\'\i}nez-Delgado}, {Miller}, {No{\"e}l}, {Sakowska},
  {Sand}, {Stringfellow}, {Tollerud}, {Vivas}, {Carballo-Bello},
  {Hernandez-Lang}, {James}, {Nidever}, {Castellon}, {Olsen}, {Zenteno}, \&
  {Delve Collaboration}}]{2023ApJ...942..111C}
{Cerny}, W., {Simon}, J.~D., {Li}, T.~S., {et~al.} 2023, \apj, 942, 111

\bibitem[{{Choi} {et~al.}(2018{\natexlab{a}}){Choi}, {Nidever}, {Olsen},
  {Blum}, {Besla}, {Zaritsky}, {van der Marel}, {Bell}, {Gallart}, {Cioni},
  {Johnson}, {Vivas}, {Saha}, {de Boer}, {No{\"e}l}, {Monachesi}, {Massana},
  {Conn}, {Martinez-Delgado}, {Mu{\~n}oz}, \&
  {Stringfellow}}]{2018ApJ...866...90C}
{Choi}, Y., {Nidever}, D.~L., {Olsen}, K., {et~al.} 2018{\natexlab{a}}, \apj,
  866, 90

\bibitem[{{Choi} {et~al.}(2018{\natexlab{b}}){Choi}, {Nidever}, {Olsen},
  {Besla}, {Blum}, {Zaritsky}, {Cioni}, {van der Marel}, {Bell}, {Johnson},
  {Vivas}, {Walker}, {de Boer}, {No{\"e}l}, {Monachesi}, {Gallart}, {Monelli},
  {Stringfellow}, {Massana}, {Martinez-Delgado}, \&
  {Mu{\~n}oz}}]{2018ApJ...869..125C}
---. 2018{\natexlab{b}}, \apj, 869, 125

\bibitem[{{Cioni} {et~al.}(2011){Cioni}, {Clementini}, {Girardi}, {Guandalini},
  {Gullieuszik}, {Miszalski}, {Moretti}, {Ripepi}, {Rubele}, {Bagheri},
  {Bekki}, {Cross}, {de Blok}, {de Grijs}, {Emerson}, {Evans}, {Gibson},
  {Gonzales-Solares}, {Groenewegen}, {Irwin}, {Ivanov}, {Lewis}, {Marconi},
  {Marquette}, {Mastropietro}, {Moore}, {Napiwotzki}, {Naylor}, {Oliveira},
  {Read}, {Sutorius}, {van Loon}, {Wilkinson}, \& {Wood}}]{2011A&A...527A.116C}
{Cioni}, M. R.~L., {Clementini}, G., {Girardi}, L., {et~al.} 2011, \aap, 527,
  A116

\bibitem[{{Conn} {et~al.}(2018){Conn}, {Jerjen}, {Kim}, \&
  {Schirmer}}]{2018ApJ...852...68C}
{Conn}, B.~C., {Jerjen}, H., {Kim}, D., \& {Schirmer}, M. 2018, \apj, 852, 68

\bibitem[{{Crnojevi{\'c}} {et~al.}(2016){Crnojevi{\'c}}, {Sand}, {Zaritsky},
  {Spekkens}, {Willman}, \& {Hargis}}]{2016ApJ...824L..14C}
{Crnojevi{\'c}}, D., {Sand}, D.~J., {Zaritsky}, D., {et~al.} 2016, \apjl, 824,
  L14

\bibitem[{{Cullinane} {et~al.}(2022){Cullinane}, {Mackey}, {Da Costa}, {Erkal},
  {Koposov}, \& {Belokurov}}]{2022MNRAS.510..445C}
{Cullinane}, L.~R., {Mackey}, A.~D., {Da Costa}, G.~S., {et~al.} 2022, \mnras,
  510, 445

\bibitem[{{DES Collaboration}(2005)}]{2005astro.ph.10346T}
{DES Collaboration}. 2005, arXiv e-prints, astro

\bibitem[{{Dias} {et~al.}(2021){Dias}, {Angelo}, {Oliveira}, {Maia}, {Parisi},
  {De Bortoli}, {Souza}, {Katime Santrich}, {Bassino}, {Barbuy}, {Bica},
  {Geisler}, {Kerber}, {P{\'e}rez-Villegas}, {Quint}, {Sanmartim}, {Santos}, \&
  {Westera}}]{2021A&A...647L...9D}
{Dias}, B., {Angelo}, M.~S., {Oliveira}, R.~A.~P., {et~al.} 2021, \aap, 647, L9

\bibitem[{{Dooley} {et~al.}(2017){Dooley}, {Peter}, {Carlin}, {Frebel},
  {Bechtol}, \& {Willman}}]{2017MNRAS.472.1060D}
{Dooley}, G.~A., {Peter}, A. H.~G., {Carlin}, J.~L., {et~al.} 2017, \mnras,
  472, 1060

\bibitem[{{Drlica-Wagner} {et~al.}(2015){Drlica-Wagner}, {Bechtol}, {Rykoff},
  {Luque}, {Queiroz}, {Mao}, {Wechsler}, {Simon}, {Santiago}, {Yanny},
  {Balbinot}, {Dodelson}, {Fausti Neto}, {James}, {Li}, {Maia}, {Marshall}, {},
  {Stringer}, {Walker}, {Abbott}, {Abdalla}, {Allam}, {Benoit-L{\'e}vy},
  {Bernstein}, {Bertin}, {Brooks}, {Buckley-Geer}, {Burke}, {Carnero Rosell},
  {Carrasco Kind}, {Carretero}, {Crocce}, {da Costa}, {Desai}, {Diehl},
  {Dietrich}, {Doel}, {Eifler}, {Evrard}, {Finley}, {Flaugher}, {Fosalba},
  {Frieman}, {Gaztanaga}, {Gerdes}, {Gruen}, {Gruendl}, {Gutierrez},
  {Honscheid}, {Kuehn}, {Kuropatkin}, {Lahav}, {Martini}, {Miquel}, {Nord},
  {Ogando}, {Plazas}, {Reil}, {Roodman}, {Sako}, {Sanchez}, {Scarpine},
  {Schubnell}, {Sevilla-Noarbe}, {Smith}, {Soares-Santos}, {Sobreira},
  {Suchyta}, {Swanson}, {Tarle}, {Tucker}, {Vikram}, {Wester}, {Zhang},
  {Zuntz}, \& {DES Collaboration}}]{2015ApJ...813..109D}
{Drlica-Wagner}, A., {Bechtol}, K., {Rykoff}, E.~S., {et~al.} 2015, \apj, 813,
  109

\bibitem[{{Drlica-Wagner} {et~al.}(2016){Drlica-Wagner}, {Bechtol}, {Allam},
  {Tucker}, {Gruendl}, {Johnson}, {Walker}, {James}, {Nidever}, {Olsen},
  {Wechsler}, {Cioni}, {Conn}, {Kuehn}, {Li}, {Mao}, {Martin}, {Neilsen},
  {Noel}, {Pieres}, {Simon}, {Stringfellow}, {van der Marel}, \&
  {Yanny}}]{2016ApJ...833L...5D}
{Drlica-Wagner}, A., {Bechtol}, K., {Allam}, S., {et~al.} 2016, \apjl, 833, L5

\bibitem[{{Drlica-Wagner} {et~al.}(2020){Drlica-Wagner}, {Bechtol}, {Mau},
  {McNanna}, {Nadler}, {Pace}, {Li}, {Pieres}, {Rozo}, {Simon}, {Walker},
  {Wechsler}, {Abbott}, {Allam}, {Annis}, {Bertin}, {Brooks}, {Burke},
  {Rosell}, {Carrasco Kind}, {Carretero}, {Costanzi}, {da Costa}, {De Vicente},
  {Desai}, {Diehl}, {Doel}, {Eifler}, {Everett}, {Flaugher}, {Frieman},
  {Garc{\'\i}a-Bellido}, {Gaztanaga}, {Gruen}, {Gruendl}, {Gschwend},
  {Gutierrez}, {Honscheid}, {James}, {Krause}, {Kuehn}, {Kuropatkin}, {Lahav},
  {Maia}, {Marshall}, {Melchior}, {Menanteau}, {Miquel}, {Palmese}, {Plazas},
  {Sanchez}, {Scarpine}, {Schubnell}, {Serrano}, {Sevilla-Noarbe}, {Smith},
  {Suchyta}, {Tarle}, \& {DES Collaboration}}]{2020ApJ...893...47D}
{Drlica-Wagner}, A., {Bechtol}, K., {Mau}, S., {et~al.} 2020, \apj, 893, 47

\bibitem[{{Drlica-Wagner} {et~al.}(2022){Drlica-Wagner}, {Ferguson},
  {Adam{\'o}w}, {Aguena}, {Allam}, {Andrade-Oliveira}, {Bacon}, {Bechtol},
  {Bell}, {Bertin}, {Bilaji}, {Bocquet}, {Bom}, {Brooks}, {Burke},
  {Carballo-Bello}, {Carlin}, {Carnero Rosell}, {Carrasco Kind}, {Carretero},
  {Castander}, {Cerny}, {Chang}, {Choi}, {Conselice}, {Costanzi},
  {Crnojevi{\'c}}, {da Costa}, {de Vicente}, {Desai}, {Esteves}, {Everett},
  {Ferrero}, {Fitzpatrick}, {Flaugher}, {Friedel}, {Frieman},
  {Garc{\'\i}a-Bellido}, {Gatti}, {Gaztanaga}, {Gerdes}, {Gruen}, {Gruendl},
  {Gschwend}, {Hartley}, {Hernandez-Lang}, {Hinton}, {Hollowood}, {Honscheid},
  {Hughes}, {Jacques}, {James}, {Johnson}, {Kuehn}, {Kuropatkin}, {Lahav},
  {Li}, {Lidman}, {Lin}, {March}, {Marshall}, {Mart{\'\i}nez-Delgado},
  {Mart{\'\i}nez-V{\'a}zquez}, {Massana}, {Mau}, {McNanna}, {Melchior},
  {Menanteau}, {Miller}, {Miquel}, {Mohr}, {Morgan}, {Mutlu-Pakdil},
  {Mu{\~n}oz}, {Neilsen}, {Nidever}, {Nikutta}, {Nilo Castellon}, {No{\"e}l},
  {Ogando}, {Olsen}, {Pace}, {Palmese}, {Paz-Chinch{\'o}n}, {Pereira},
  {Pieres}, {Plazas Malag{\'o}n}, {Prat}, {Riley}, {Rodriguez-Monroy}, {Romer},
  {Roodman}, {Sako}, {Sakowska}, {Sanchez}, {S{\'a}nchez}, {Sand},
  {Santana-Silva}, {Santiago}, {Schubnell}, {Serrano}, {Sevilla-Noarbe},
  {Simon}, {Smith}, {Soares-Santos}, {Stringfellow}, {Suchyta}, {Suson}, {Tan},
  {Tarle}, {Tavangar}, {Thomas}, {To}, {Tollerud}, {Troxel}, {Tucker}, {Varga},
  {Vivas}, {Walker}, {Weller}, {Wilkinson}, {Wu}, {Yanny}, {Zaborowski},
  {Zenteno}, {Delve Collaboration}, {Des Collaboration}, \& {Astro Data
  Lab}}]{2022ApJS..261...38D}
{Drlica-Wagner}, A., {Ferguson}, P.~S., {Adam{\'o}w}, M., {et~al.} 2022, \apjs,
  261, 38

\bibitem[{{El Youssoufi} {et~al.}(2021){El Youssoufi}, {Cioni}, {Bell}, {de
  Grijs}, {Groenewegen}, {Ivanov}, {Matijev{\u{i}}c}, {Niederhofer},
  {Oliveira}, {Ripepi}, {Schmidt}, {Subramanian}, {Sun}, \& {van
  Loon}}]{2021MNRAS.505.2020E}
{El Youssoufi}, D., {Cioni}, M.-R.~L., {Bell}, C. P.~M., {et~al.} 2021, \mnras,
  505, 2020

\bibitem[{{Erkal} \& {Belokurov}(2020)}]{2020MNRAS.495.2554E}
{Erkal}, D., \& {Belokurov}, V.~A. 2020, \mnras, 495, 2554

\bibitem[{{Fadely} {et~al.}(2011){Fadely}, {Willman}, {Geha}, {Walsh},
  {Mu{\~n}oz}, {Jerjen}, {Vargas}, \& {Da Costa}}]{2011AJ....142...88F}
{Fadely}, R., {Willman}, B., {Geha}, M., {et~al.} 2011, \aj, 142, 88

\bibitem[{{Flaugher} {et~al.}(2015){Flaugher}, {Diehl}, {Honscheid}, {Abbott},
  {Alvarez}, {Angstadt}, {Annis}, {Antonik}, {Ballester}, {Beaufore},
  {Bernstein}, {Bernstein}, {Bigelow}, {Bonati}, {Boprie}, {Brooks},
  {Buckley-Geer}, {Campa}, {Cardiel-Sas}, {Castander}, {Castilla}, {Cease},
  {Cela-Ruiz}, {Chappa}, {Chi}, {Cooper}, {da Costa}, {Dede}, {Derylo},
  {DePoy}, {de Vicente}, {Doel}, {Drlica-Wagner}, {Eiting}, {Elliott}, {Emes},
  {Estrada}, {Fausti Neto}, {Finley}, {Flores}, {Frieman}, {Gerdes},
  {Gladders}, {Gregory}, {Gutierrez}, {Hao}, {Holland}, {Holm}, {Huffman},
  {Jackson}, {James}, {Jonas}, {Karcher}, {Karliner}, {Kent}, {Kessler},
  {Kozlovsky}, {Kron}, {Kubik}, {Kuehn}, {Kuhlmann}, {Kuk}, {Lahav}, {Lathrop},
  {Lee}, {Levi}, {Lewis}, {Li}, {Mandrichenko}, {Marshall}, {Martinez},
  {Merritt}, {Miquel}, {Mu{\~n}oz}, {Neilsen}, {Nichol}, {Nord}, {Ogando},
  {Olsen}, {Palaio}, {Patton}, {Peoples}, {Plazas}, {Rauch}, {Reil}, {Rheault},
  {Roe}, {Rogers}, {Roodman}, {Sanchez}, {Scarpine}, {Schindler}, {Schmidt},
  {Schmitt}, {Schubnell}, {Schultz}, {Schurter}, {Scott}, {Serrano}, {Shaw},
  {Smith}, {Soares-Santos}, {Stefanik}, {Stuermer}, {Suchyta}, {Sypniewski},
  {Tarle}, {Thaler}, {Tighe}, {Tran}, {Tucker}, {Walker}, {Wang}, {Watson},
  {Weaverdyck}, {Wester}, {Woods}, {Yanny}, \& {DES
  Collaboration}}]{2015AJ....150..150F}
{Flaugher}, B., {Diehl}, H.~T., {Honscheid}, K., {et~al.} 2015, \aj, 150, 150

\bibitem[{{Foreman-Mackey} {et~al.}(2013){Foreman-Mackey}, {Hogg}, {Lang}, \&
  {Goodman}}]{Foreman-Mackey:2013}
{Foreman-Mackey}, D., {Hogg}, D.~W., {Lang}, D., \& {Goodman}, J. 2013, \pasp,
  125, 306

\bibitem[{{Gaia Collaboration} {et~al.}(2016){Gaia Collaboration}, {Prusti},
  {de Bruijne}, {Brown}, {Vallenari}, {Babusiaux}, {Bailer-Jones}, {Bastian},
  {Biermann}, {Evans}, {Eyer}, {Jansen}, {Jordi}, {Klioner}, {Lammers},
  {Lindegren}, {Luri}, {Mignard}, {Milligan}, {Panem}, {Poinsignon},
  {Pourbaix}, {Randich}, {Sarri}, {Sartoretti}, {Siddiqui}, {Soubiran},
  {Valette}, {van Leeuwen}, {Walton}, {Aerts}, {Arenou}, {Cropper}, {Drimmel},
  {H{\o}g}, {Katz}, {Lattanzi}, {O'Mullane}, {Grebel}, {Holland}, {Huc},
  {Passot}, {Bramante}, {Cacciari}, {Casta{\~n}eda}, {Chaoul}, {Cheek}, {De
  Angeli}, {Fabricius}, {Guerra}, {Hern{\'a}ndez}, {Jean-Antoine-Piccolo},
  {Masana}, {Messineo}, {Mowlavi}, {Nienartowicz}, {Ord{\'o}{\~n}ez-Blanco},
  {Panuzzo}, {Portell}, {Richards}, {Riello}, {Seabroke}, {Tanga},
  {Th{\'e}venin}, {Torra}, {Els}, {Gracia-Abril}, {Comoretto},
  {Garcia-Reinaldos}, {Lock}, {Mercier}, {Altmann}, {Andrae}, {Astraatmadja},
  {Bellas-Velidis}, {Benson}, {Berthier}, {Blomme}, {Busso}, {Carry},
  {Cellino}, {Clementini}, {Cowell}, {Creevey}, {Cuypers}, {Davidson}, {De
  Ridder}, {de Torres}, {Delchambre}, {Dell'Oro}, {Ducourant}, {Fr{\'e}mat},
  {Garc{\'\i}a-Torres}, {Gosset}, {Halbwachs}, {Hambly}, {Harrison}, {Hauser},
  {Hestroffer}, {Hodgkin}, {Huckle}, {Hutton}, {Jasniewicz}, {Jordan},
  {Kontizas}, {Korn}, {Lanzafame}, {Manteiga}, {Moitinho}, {Muinonen},
  {Osinde}, {Pancino}, {Pauwels}, {Petit}, {Recio-Blanco}, {Robin}, {Sarro},
  {Siopis}, {Smith}, {Smith}, {Sozzetti}, {Thuillot}, {van Reeven}, {Viala},
  {Abbas}, {Abreu Aramburu}, {Accart}, {Aguado}, {Allan}, {Allasia},
  {Altavilla}, {{\'A}lvarez}, {Alves}, {Anderson}, {Andrei}, {Anglada Varela},
  {Antiche}, {Antoja}, {Ant{\'o}n}, {Arcay}, {Atzei}, {Ayache}, {Bach},
  {Baker}, {Balaguer-N{\'u}{\~n}ez}, {Barache}, {Barata}, {Barbier}, {Barblan},
  {Baroni}, {Barrado y Navascu{\'e}s}, {Barros}, {Barstow}, {Becciani},
  {Bellazzini}, {Bellei}, {Bello Garc{\'\i}a}, {Belokurov}, {Bendjoya},
  {Berihuete}, {Bianchi}, {Bienaym{\'e}}, {Billebaud}, {Blagorodnova},
  {Blanco-Cuaresma}, {Boch}, {Bombrun}, {Borrachero}, {Bouquillon}, {Bourda},
  {Bouy}, {Bragaglia}, {Breddels}, {Brouillet}, {Br{\"u}semeister},
  {Bucciarelli}, {Budnik}, {Burgess}, {Burgon}, {Burlacu}, {Busonero}, {Buzzi},
  {Caffau}, {Cambras}, {Campbell}, {Cancelliere}, {Cantat-Gaudin}, {Carlucci},
  {Carrasco}, {Castellani}, {Charlot}, {Charnas}, {Charvet}, {Chassat},
  {Chiavassa}, {Clotet}, {Cocozza}, {Collins}, {Collins}, {Costigan}, {Crifo},
  {Cross}, {Crosta}, {Crowley}, {Dafonte}, {Damerdji}, {Dapergolas}, {David},
  {David}, {De Cat}, {de Felice}, {de Laverny}, {De Luise}, {De March}, {de
  Martino}, {de Souza}, {Debosscher}, {del Pozo}, {Delbo}, {Delgado},
  {Delgado}, {di Marco}, {Di Matteo}, {Diakite}, {Distefano}, {Dolding}, {Dos
  Anjos}, {Drazinos}, {Dur{\'a}n}, {Dzigan}, {Ecale}, {Edvardsson}, {Enke},
  {Erdmann}, {Escolar}, {Espina}, {Evans}, {Eynard Bontemps}, {Fabre},
  {Fabrizio}, {Faigler}, {Falc{\~a}o}, {Farr{\`a}s Casas}, {Faye}, {Federici},
  {Fedorets}, {Fern{\'a}ndez-Hern{\'a}ndez}, {Fernique}, {Fienga}, {Figueras},
  {Filippi}, {Findeisen}, {Fonti}, {Fouesneau}, {Fraile}, {Fraser}, {Fuchs},
  {Furnell}, {Gai}, {Galleti}, {Galluccio}, {Garabato}, {Garc{\'\i}a-Sedano},
  {Gar{\'e}}, {Garofalo}, {Garralda}, {Gavras}, {Gerssen}, {Geyer}, {Gilmore},
  {Girona}, {Giuffrida}, {Gomes}, {Gonz{\'a}lez-Marcos},
  {Gonz{\'a}lez-N{\'u}{\~n}ez}, {Gonz{\'a}lez-Vidal}, {Granvik}, {Guerrier},
  {Guillout}, {Guiraud}, {G{\'u}rpide}, {Guti{\'e}rrez-S{\'a}nchez}, {Guy},
  {Haigron}, {Hatzidimitriou}, {Haywood}, {Heiter}, {Helmi}, {Hobbs},
  {Hofmann}, {Holl}, {Holland}, {Hunt}, {Hypki}, {Icardi}, {Irwin}, {Jevardat
  de Fombelle}, {Jofr{\'e}}, {Jonker}, {Jorissen}, {Julbe}, {Karampelas},
  {Kochoska}, {Kohley}, {Kolenberg}, {Kontizas}, {Koposov}, {Kordopatis},
  {Koubsky}, {Kowalczyk}, {Krone-Martins}, {Kudryashova}, {Kull}, {Bachchan},
  {Lacoste-Seris}, {Lanza}, {Lavigne}, {Le Poncin-Lafitte}, {Lebreton},
  {Lebzelter}, {Leccia}, {Leclerc}, {Lecoeur-Taibi}, {Lemaitre}, {Lenhardt},
  {Leroux}, {Liao}, {Licata}, {Lindstr{\o}m}, {Lister}, {Livanou}, {Lobel},
  {L{\"o}ffler}, {L{\'o}pez}, {Lopez-Lozano}, {Lorenz}, {Loureiro},
  {MacDonald}, {Magalh{\~a}es Fernandes}, {Managau}, {Mann}, {Mantelet},
  {Marchal}, {Marchant}, {Marconi}, {Marie}, {Marinoni}, {Marrese},
  {Marschalk{\'o}}, {Marshall}, {Mart{\'\i}n-Fleitas}, {Martino}, {Mary},
  {Matijevi{\v{c}}}, {Mazeh}, {McMillan}, {Messina}, {Mestre}, {Michalik},
  {Millar}, {Miranda}, {Molina}, {Molinaro}, {Molinaro}, {Moln{\'a}r},
  {Moniez}, {Montegriffo}, {Monteiro}, {Mor}, {Mora}, {Morbidelli}, {Morel},
  {Morgenthaler}, {Morley}, {Morris}, {Mulone}, {Muraveva}, {Musella},
  {Narbonne}, {Nelemans}, {Nicastro}, {Noval}, {Ord{\'e}novic},
  {Ordieres-Mer{\'e}}, {Osborne}, {Pagani}, {Pagano}, {Pailler}, {Palacin},
  {Palaversa}, {Parsons}, {Paulsen}, {Pecoraro}, {Pedrosa}, {Pentik{\"a}inen},
  {Pereira}, {Pichon}, {Piersimoni}, {Pineau}, {Plachy}, {Plum}, {Poujoulet},
  {Pr{\v{s}}a}, {Pulone}, {Ragaini}, {Rago}, {Rambaux}, {Ramos-Lerate},
  {Ranalli}, {Rauw}, {Read}, {Regibo}, {Renk}, {Reyl{\'e}}, {Ribeiro},
  {Rimoldini}, {Ripepi}, {Riva}, {Rixon}, {Roelens}, {Romero-G{\'o}mez},
  {Rowell}, {Royer}, {Rudolph}, {Ruiz-Dern}, {Sadowski}, {Sagrist{\`a}
  Sell{\'e}s}, {Sahlmann}, {Salgado}, {Salguero}, {Sarasso}, {Savietto},
  {Schnorhk}, {Schultheis}, {Sciacca}, {Segol}, {Segovia}, {Segransan},
  {Serpell}, {Shih}, {Smareglia}, {Smart}, {Smith}, {Solano}, {Solitro},
  {Sordo}, {Soria Nieto}, {Souchay}, {Spagna}, {Spoto}, {Stampa}, {Steele},
  {Steidelm{\"u}ller}, {Stephenson}, {Stoev}, {Suess}, {S{\"u}veges}, {Surdej},
  {Szabados}, {Szegedi-Elek}, {Tapiador}, {Taris}, {Tauran}, {Taylor},
  {Teixeira}, {Terrett}, {Tingley}, {Trager}, {Turon}, {Ulla}, {Utrilla},
  {Valentini}, {van Elteren}, {Van Hemelryck}, {van Leeuwen}, {Varadi},
  {Vecchiato}, {Veljanoski}, {Via}, {Vicente}, {Vogt}, {Voss}, {Votruba},
  {Voutsinas}, {Walmsley}, {Weiler}, {Weingrill}, {Werner}, {Wevers},
  {Whitehead}, {Wyrzykowski}, {Yoldas}, {{\v{Z}}erjal}, {Zucker}, {Zurbach},
  {Zwitter}, {Alecu}, {Allen}, {Allende Prieto}, {Amorim},
  {Anglada-Escud{\'e}}, {Arsenijevic}, {Azaz}, {Balm}, {Beck}, {Bernstein},
  {Bigot}, {Bijaoui}, {Blasco}, {Bonfigli}, {Bono}, {Boudreault}, {Bressan},
  {Brown}, {Brunet}, {Bunclark}, {Buonanno}, {Butkevich}, {Carret}, {Carrion},
  {Chemin}, {Ch{\'e}reau}, {Corcione}, {Darmigny}, {de Boer}, {de Teodoro}, {de
  Zeeuw}, {Delle Luche}, {Domingues}, {Dubath}, {Fodor}, {Fr{\'e}zouls},
  {Fries}, {Fustes}, {Fyfe}, {Gallardo}, {Gallegos}, {Gardiol}, {Gebran},
  {Gomboc}, {G{\'o}mez}, {Grux}, {Gueguen}, {Heyrovsky}, {Hoar}, {Iannicola},
  {Isasi Parache}, {Janotto}, {Joliet}, {Jonckheere}, {Keil}, {Kim},
  {Klagyivik}, {Klar}, {Knude}, {Kochukhov}, {Kolka}, {Kos}, {Kutka}, {Lainey},
  {LeBouquin}, {Liu}, {Loreggia}, {Makarov}, {Marseille}, {Martayan},
  {Martinez-Rubi}, {Massart}, {Meynadier}, {Mignot}, {Munari}, {Nguyen},
  {Nordlander}, {Ocvirk}, {O'Flaherty}, {Olias Sanz}, {Ortiz}, {Osorio},
  {Oszkiewicz}, {Ouzounis}, {Palmer}, {Park}, {Pasquato}, {Peltzer}, {Peralta},
  {P{\'e}turaud}, {Pieniluoma}, {Pigozzi}, {Poels}, {Prat}, {Prod'homme},
  {Raison}, {Rebordao}, {Risquez}, {Rocca-Volmerange}, {Rosen}, {Ruiz-Fuertes},
  {Russo}, {Sembay}, {Serraller Vizcaino}, {Short}, {Siebert}, {Silva},
  {Sinachopoulos}, {Slezak}, {Soffel}, {Sosnowska}, {Strai{\v{z}}ys}, {ter
  Linden}, {Terrell}, {Theil}, {Tiede}, {Troisi}, {Tsalmantza}, {Tur},
  {Vaccari}, {Vachier}, {Valles}, {Van Hamme}, {Veltz}, {Virtanen}, {Wallut},
  {Wichmann}, {Wilkinson}, {Ziaeepour}, \& {Zschocke}}]{2016A&A...595A...1G}
{Gaia Collaboration}, {Prusti}, T., {de Bruijne}, J.~H.~J., {et~al.} 2016,
  \aap, 595, A1

\bibitem[{{Gaia Collaboration} {et~al.}(2022){Gaia Collaboration}, {Vallenari},
  {Brown}, {Prusti}, {de Bruijne}, {Arenou}, {Babusiaux}, {Biermann},
  {Creevey}, {Ducourant}, {Evans}, {Eyer}, {Guerra}, {Hutton}, {Jordi},
  {Klioner}, {Lammers}, {Lindegren}, {Luri}, {Mignard}, {Panem}, {Pourbaix},
  {Randich}, {Sartoretti}, {Soubiran}, {Tanga}, {Walton}, {Bailer-Jones},
  {Bastian}, {Drimmel}, {Jansen}, {Katz}, {Lattanzi}, {van Leeuwen}, {Bakker},
  {Cacciari}, {Casta{\~n}eda}, {De Angeli}, {Fabricius}, {Fouesneau},
  {Fr{\'e}mat}, {Galluccio}, {Guerrier}, {Heiter}, {Masana}, {Messineo},
  {Mowlavi}, {Nicolas}, {Nienartowicz}, {Pailler}, {Panuzzo}, {Riclet}, {Roux},
  {Seabroke}, {Sordo{\o}rcit}, {Th{\'e}venin}, {Gracia-Abril}, {Portell},
  {Teyssier}, {Altmann}, {Andrae}, {Audard}, {Bellas-Velidis}, {Benson},
  {Berthier}, {Blomme}, {Burgess}, {Busonero}, {Busso}, {C{\'a}novas}, {Carry},
  {Cellino}, {Cheek}, {Clementini}, {Damerdji}, {Davidson}, {de Teodoro},
  {Nu{\~n}ez Campos}, {Delchambre}, {Dell'Oro}, {Esquej},
  {Fern{\'a}ndez-Hern{\'a}ndez}, {Fraile}, {Garabato}, {Garc{\'\i}a-Lario},
  {Gosset}, {Haigron}, {Halbwachs}, {Hambly}, {Harrison}, {Hern{\'a}ndez},
  {Hestroffer}, {Hodgkin}, {Holl}, {Jan{\ss}en}, {Jevardat de Fombelle},
  {Jordan}, {Krone-Martins}, {Lanzafame}, {L{\"o}ffler}, {Marchal}, {Marrese},
  {Moitinho}, {Muinonen}, {Osborne}, {Pancino}, {Pauwels}, {Recio-Blanco},
  {Reyl{\'e}}, {Riello}, {Rimoldini}, {Roegiers}, {Rybizki}, {Sarro}, {Siopis},
  {Smith}, {Sozzetti}, {Utrilla}, {van Leeuwen}, {Abbas}, {{\'A}brah{\'a}m},
  {Abreu Aramburu}, {Aerts}, {Aguado}, {Ajaj}, {Aldea-Montero}, {Altavilla},
  {{\'A}lvarez}, {Alves}, {Anders}, {Anderson}, {Anglada Varela}, {Antoja},
  {Baines}, {Baker}, {Balaguer-N{\'u}{\~n}ez}, {Balbinot}, {Balog}, {Barache},
  {Barbato}, {Barros}, {Barstow}, {Bartolom{\'e}}, {Bassilana}, {Bauchet},
  {Becciani}, {Bellazzini}, {Berihuete}, {Bernet}, {Bertone}, {Bianchi},
  {Binnenfeld}, {Blanco-Cuaresma}, {Blazere}, {Boch}, {Bombrun}, {Bossini},
  {Bouquillon}, {Bragaglia}, {Bramante}, {Breedt}, {Bressan}, {Brouillet},
  {Brugaletta}, {Bucciarelli}, {Burlacu}, {Butkevich}, {Buzzi}, {Caffau},
  {Cancelliere}, {Cantat-Gaudin}, {Carballo}, {Carlucci}, {Carnerero},
  {Carrasco}, {Casamiquela}, {Castellani}, {Castro-Ginard}, {Chaoul},
  {Charlot}, {Chemin}, {Chiaramida}, {Chiavassa}, {Chornay}, {Comoretto},
  {Contursi}, {Cooper}, {Cornez}, {Cowell}, {Crifo}, {Cropper}, {Crosta},
  {Crowley}, {Dafonte}, {Dapergolas}, {David}, {David}, {de Laverny}, {De
  Luise}, {De March}, {De Ridder}, {de Souza}, {de Torres}, {del Peloso}, {del
  Pozo}, {Delbo}, {Delgado}, {Delisle}, {Demouchy}, {Dharmawardena}, {Di
  Matteo}, {Diakite}, {Diener}, {Distefano}, {Dolding}, {Edvardsson}, {Enke},
  {Fabre}, {Fabrizio}, {Faigler}, {Fedorets}, {Fernique}, {Fienga}, {Figueras},
  {Fournier}, {Fouron}, {Fragkoudi}, {Gai}, {Garcia-Gutierrez},
  {Garcia-Reinaldos}, {Garc{\'\i}a-Torres}, {Garofalo}, {Gavel}, {Gavras},
  {Gerlach}, {Geyer}, {Giacobbe}, {Gilmore}, {Girona}, {Giuffrida}, {Gomel},
  {Gomez}, {Gonz{\'a}lez-N{\'u}{\~n}ez}, {Gonz{\'a}lez-Santamar{\'\i}a},
  {Gonz{\'a}lez-Vidal}, {Granvik}, {Guillout}, {Guiraud},
  {Guti{\'e}rrez-S{\'a}nchez}, {Guy}, {Hatzidimitriou}, {Hauser}, {Haywood},
  {Helmer}, {Helmi}, {Sarmiento}, {Hidalgo}, {Hilger}, {H{\l}adczuk}, {Hobbs},
  {Holland}, {Huckle}, {Jardine}, {Jasniewicz}, {Jean-Antoine Piccolo},
  {Jim{\'e}nez-Arranz}, {Jorissen}, {Juaristi Campillo}, {Julbe}, {Karbevska},
  {Kervella}, {Khanna}, {Kontizas}, {Kordopatis}, {Korn}, {K{\'o}sp{\'a}l},
  {Kostrzewa-Rutkowska}, {Kruszy{\'n}ska}, {Kun}, {Laizeau}, {Lambert},
  {Lanza}, {Lasne}, {Le Campion}, {Lebreton}, {Lebzelter}, {Leccia}, {Leclerc},
  {Lecoeur-Taibi}, {Liao}, {Licata}, {Lindstr{\o}m}, {Lister}, {Livanou},
  {Lobel}, {Lorca}, {Loup}, {Madrero Pardo}, {Magdaleno Romeo}, {Managau},
  {Mann}, {Manteiga}, {Marchant}, {Marconi}, {Marcos}, {Marcos Santos},
  {Mar{\'\i}n Pina}, {Marinoni}, {Marocco}, {Marshall}, {Polo},
  {Mart{\'\i}n-Fleitas}, {Marton}, {Mary}, {Masip}, {Massari},
  {Mastrobuono-Battisti}, {Mazeh}, {McMillan}, {Messina}, {Michalik}, {Millar},
  {Mints}, {Molina}, {Molinaro}, {Moln{\'a}r}, {Monari}, {Mongui{\'o}},
  {Montegriffo}, {Montero}, {Mor}, {Mora}, {Morbidelli}, {Morel}, {Morris},
  {Muraveva}, {Murphy}, {Musella}, {Nagy}, {Noval}, {Oca{\~n}a}, {Ogden},
  {Ordenovic}, {Osinde}, {Pagani}, {Pagano}, {Palaversa}, {Palicio},
  {Pallas-Quintela}, {Panahi}, {Payne-Wardenaar}, {Pe{\~n}alosa Esteller},
  {Penttil{\"a}}, {Pichon}, {Piersimoni}, {Pineau}, {Plachy}, {Plum}, {Poggio},
  {Pr{\v{s}}a}, {Pulone}, {Racero}, {Ragaini}, {Rainer}, {Raiteri}, {Rambaux},
  {Ramos}, {Ramos-Lerate}, {Re Fiorentin}, {Regibo}, {Richards}, {Rios Diaz},
  {Ripepi}, {Riva}, {Rix}, {Rixon}, {Robichon}, {Robin}, {Robin}, {Roelens},
  {Rogues}, {Rohrbasser}, {Romero-G{\'o}mez}, {Rowell}, {Royer}, {Ruz Mieres},
  {Rybicki}, {Sadowski}, {S{\'a}ez N{\'u}{\~n}ez}, {Sagrist{\`a} Sell{\'e}s},
  {Sahlmann}, {Salguero}, {Samaras}, {Sanchez Gimenez}, {Sanna},
  {Santove{\~n}a}, {Sarasso}, {Schultheis}, {Sciacca}, {Segol}, {Segovia},
  {S{\'e}gransan}, {Semeux}, {Shahaf}, {Siddiqui}, {Siebert}, {Siltala},
  {Silvelo}, {Slezak}, {Slezak}, {Smart}, {Snaith}, {Solano}, {Solitro},
  {Souami}, {Souchay}, {Spagna}, {Spina}, {Spoto}, {Steele},
  {Steidelm{\"u}ller}, {Stephenson}, {S{\"u}veges}, {Surdej}, {Szabados},
  {Szegedi-Elek}, {Taris}, {Taylo}, {Teixeira}, {Tolomei}, {Tonello}, {Torra},
  {Torra}, {Torralba Elipe}, {Trabucchi}, {Tsounis}, {Turon}, {Ulla}, {Unger},
  {Vaillant}, {van Dillen}, {van Reeven}, {Vanel}, {Vecchiato}, {Viala},
  {Vicente}, {Voutsinas}, {Weiler}, {Wevers}, {Wyrzykowski}, {Yoldas}, {Yvard},
  {Zhao}, {Zorec}, {Zucker}, \& {Zwitter}}]{2022arXiv220800211G}
{Gaia Collaboration}, {Vallenari}, A., {Brown}, A.~G.~A., {et~al.} 2022, arXiv
  e-prints, arXiv:2208.00211

\bibitem[{{Garavito-Camargo} {et~al.}(2021){Garavito-Camargo}, {Besla},
  {Laporte}, {Price-Whelan}, {Cunningham}, {Johnston}, {Weinberg}, \&
  {G{\'o}mez}}]{2021ApJ...919..109G}
{Garavito-Camargo}, N., {Besla}, G., {Laporte}, C. F.~P., {et~al.} 2021, \apj,
  919, 109

\bibitem[{{Gatto} {et~al.}(2022{\natexlab{a}}){Gatto}, {Ripepi}, {Bellazzini},
  {Tortora}, {Tosi}, {Cignoni}, \& {Longo}}]{2022ApJ...931...19G}
{Gatto}, M., {Ripepi}, V., {Bellazzini}, M., {et~al.} 2022{\natexlab{a}}, \apj,
  931, 19

\bibitem[{{Gatto} {et~al.}(2020){Gatto}, {Ripepi}, {Bellazzini}, {Cignoni},
  {Cioni}, {Dall'Ora}, {Longo}, {Marconi}, {Schipani}, \&
  {Tosi}}]{2020MNRAS.499.4114G}
---. 2020, \mnras, 499, 4114

\bibitem[{{Gatto} {et~al.}(2021){Gatto}, {Ripepi}, {Bellazzini}, {Tosi},
  {Tortora}, {Cignoni}, {Spavone}, {Dall'ora}, {Clementini}, {Cusano}, {Longo},
  {Musella}, {Marconi}, \& {Schipani}}]{2021RNAAS...5..159G}
---. 2021, Research Notes of the American Astronomical Society, 5, 159

\bibitem[{{Gatto} {et~al.}(2022{\natexlab{b}}){Gatto}, {Ripepi}, {Bellazzini},
  {Dall'ora}, {Tosi}, {Tortora}, {Cignoni}, {Cioni}, {Cusano}, {Longo},
  {Marconi}, {Musella}, {Schipani}, \& {Spavone}}]{2022ApJ...929L..21G}
---. 2022{\natexlab{b}}, \apjl, 929, L21

\bibitem[{{Glatt} {et~al.}(2008){Glatt}, {Gallagher}, {Grebel}, {Nota},
  {Sabbi}, {Sirianni}, {Clementini}, {Tosi}, {Harbeck}, {Koch}, \&
  {Cracraft}}]{2008AJ....135.1106G}
{Glatt}, K., {Gallagher}, John~S., I., {Grebel}, E.~K., {et~al.} 2008, \aj,
  135, 1106

\bibitem[{{Goodman} \& {Weare}(2010)}]{2010CAMCS...5...65G}
{Goodman}, J., \& {Weare}, J. 2010, Communications in Applied Mathematics and
  Computational Science, 5, 65

\bibitem[{{G{\'o}rski} {et~al.}(2005){G{\'o}rski}, {Hivon}, {Banday},
  {Wandelt}, {Hansen}, {Reinecke}, \& {Bartelmann}}]{2005ApJ...622..759G}
{G{\'o}rski}, K.~M., {Hivon}, E., {Banday}, A.~J., {et~al.} 2005, \apj, 622,
  759

\bibitem[{{Graczyk} {et~al.}(2020){Graczyk}, {Pietrzy{\'n}ski}, {Thompson},
  {Gieren}, {Zgirski}, {Villanova}, {G{\'o}rski}, {Wielg{\'o}rski},
  {Karczmarek}, {Narloch}, {Pilecki}, {Taormina}, {Smolec}, {Suchomska},
  {Gallenne}, {Nardetto}, {Storm}, {Kudritzki}, {Ka{\l}uszy{\'n}ski}, \&
  {Pych}}]{2020ApJ...904...13G}
{Graczyk}, D., {Pietrzy{\'n}ski}, G., {Thompson}, I.~B., {et~al.} 2020, \apj,
  904, 13

\bibitem[{{GRAVITY Collaboration} {et~al.}(2019){GRAVITY Collaboration},
  {Abuter}, {Amorim}, {Baub{\"o}ck}, {Berger}, {Bonnet}, {Brandner},
  {Cl{\'e}net}, {Coud{\'e} Du Foresto}, {de Zeeuw}, {Dexter}, {Duvert},
  {Eckart}, {Eisenhauer}, {F{\"o}rster Schreiber}, {Garcia}, {Gao}, {Gendron},
  {Genzel}, {Gerhard}, {Gillessen}, {Habibi}, {Haubois}, {Henning}, {Hippler},
  {Horrobin}, {Jim{\'e}nez-Rosales}, {Jocou}, {Kervella}, {Lacour},
  {Lapeyr{\`e}re}, {Le Bouquin}, {L{\'e}na}, {Ott}, {Paumard}, {Perraut},
  {Perrin}, {Pfuhl}, {Rabien}, {Rodriguez Coira}, {Rousset}, {Scheithauer},
  {Sternberg}, {Straub}, {Straubmeier}, {Sturm}, {Tacconi}, {Vincent}, {von
  Fellenberg}, {Waisberg}, {Widmann}, {Wieprecht}, {Wiezorrek}, {Woillez}, \&
  {Yazici}}]{2019A&A...625L..10G}
{GRAVITY Collaboration}, {Abuter}, R., {Amorim}, A., {et~al.} 2019, \aap, 625,
  L10

\bibitem[{{Harris} {et~al.}(2020){Harris}, {Millman}, {van der Walt},
  {Gommers}, {Virtanen}, {Cournapeau}, {Wieser}, {Taylor}, {Berg}, {Smith},
  {Kern}, {Picus}, {Hoyer}, {van Kerkwijk}, {Brett}, {Haldane}, {del R{\'\i}o},
  {Wiebe}, {Peterson}, {G{\'e}rard-Marchant}, {Sheppard}, {Reddy}, {Weckesser},
  {Abbasi}, {Gohlke}, \& {Oliphant}}]{2020Natur.585..357H}
{Harris}, C.~R., {Millman}, K.~J., {van der Walt}, S.~J., {et~al.} 2020, \nat,
  585, 357

\bibitem[{{Harris}(2010)}]{2010arXiv1012.3224H}
{Harris}, W.~E. 2010, arXiv e-prints, arXiv:1012.3224

\bibitem[{{Hinton}(2019)}]{2019ascl.soft10017H}
{Hinton}, S.~R. 2019, {ChainConsumer: Corner plots, LaTeX tables and plotting
  walks}, Astrophysics Source Code Library, record ascl:1910.017, , ,
  ascl:1910.017

\bibitem[{{Homma} {et~al.}(2018){Homma}, {Chiba}, {Okamoto}, {Komiyama},
  {Tanaka}, {Tanaka}, {Ishigaki}, {Hayashi}, {Arimoto}, {Garmilla}, {Lupton},
  {Strauss}, {Miyazaki}, {Wang}, \& {Murayama}}]{2018PASJ...70S..18H}
{Homma}, D., {Chiba}, M., {Okamoto}, S., {et~al.} 2018, \pasj, 70, S18

\bibitem[{{Homma} {et~al.}(2019){Homma}, {Chiba}, {Komiyama}, {Tanaka},
  {Okamoto}, {Tanaka}, {Ishigaki}, {Hayashi}, {Arimoto}, {Carlsten}, {Lupton},
  {Strauss}, {Miyazaki}, {Torrealba}, {Wang}, \&
  {Murayama}}]{2019PASJ...71...94H}
{Homma}, D., {Chiba}, M., {Komiyama}, Y., {et~al.} 2019, \pasj, 71, 94

\bibitem[{Jahn {et~al.}(2019)Jahn, Sales, Wetzel, Boylan-Kolchin, Chan,
  El-Badry, Lazar, \& Bullock}]{10.1093/mnras/stz2457}
Jahn, E.~D., Sales, L.~V., Wetzel, A., {et~al.} 2019, Monthly Notices of the
  Royal Astronomical Society, 489, 5348.
\newblock \url{https://doi.org/10.1093/mnras/stz2457}

\bibitem[{{Jethwa} {et~al.}(2016){Jethwa}, {Erkal}, \&
  {Belokurov}}]{2016MNRAS.461.2212J}
{Jethwa}, P., {Erkal}, D., \& {Belokurov}, V. 2016, \mnras, 461, 2212

\bibitem[{{Ji} {et~al.}(2021){Ji}, {Koposov}, {Li}, {Erkal}, {Pace}, {Simon},
  {Belokurov}, {Cullinane}, {Da Costa}, {Kuehn}, {Lewis}, {Mackey}, {Shipp},
  {Simpson}, {Zucker}, {Hansen}, {Bland-Hawthorn}, \& {S5
  Collaboration}}]{2021ApJ...921...32J}
{Ji}, A.~P., {Koposov}, S.~E., {Li}, T.~S., {et~al.} 2021, \apj, 921, 32

\bibitem[{{Kallivayalil} {et~al.}(2018){Kallivayalil}, {Sales}, {Zivick},
  {Fritz}, {Del Pino}, {Sohn}, {Besla}, {van der Marel}, {Navarro}, \&
  {Sacchi}}]{2018ApJ...867...19K}
{Kallivayalil}, N., {Sales}, L.~V., {Zivick}, P., {et~al.} 2018, \apj, 867, 19

\bibitem[{{Kim} \& {Jerjen}(2015{\natexlab{a}})}]{2015ApJ...808L..39K}
{Kim}, D., \& {Jerjen}, H. 2015{\natexlab{a}}, \apjl, 808, L39

\bibitem[{{Kim} \& {Jerjen}(2015{\natexlab{b}})}]{2015ApJ...799...73K}
---. 2015{\natexlab{b}}, \apj, 799, 73

\bibitem[{{Kim} {et~al.}(2016){Kim}, {Jerjen}, {Mackey}, {Da Costa}, \&
  {Milone}}]{2016ApJ...820..119K}
{Kim}, D., {Jerjen}, H., {Mackey}, D., {Da Costa}, G.~S., \& {Milone}, A.~P.
  2016, \apj, 820, 119

\bibitem[{{Kim} {et~al.}(2015){Kim}, {Jerjen}, {Milone}, {Mackey}, \& {Da
  Costa}}]{2015ApJ...803...63K}
{Kim}, D., {Jerjen}, H., {Milone}, A.~P., {Mackey}, D., \& {Da Costa}, G.~S.
  2015, \apj, 803, 63

\bibitem[{{Koposov} {et~al.}(2007){Koposov}, {de Jong}, {Belokurov}, {Rix},
  {Zucker}, {Evans}, {Gilmore}, {Irwin}, \& {Bell}}]{2007ApJ...669..337K}
{Koposov}, S., {de Jong}, J.~T.~A., {Belokurov}, V., {et~al.} 2007, \apj, 669,
  337

\bibitem[{{Koposov} {et~al.}(2015){Koposov}, {Belokurov}, {Torrealba}, \&
  {Evans}}]{2015ApJ...805..130K}
{Koposov}, S.~E., {Belokurov}, V., {Torrealba}, G., \& {Evans}, N.~W. 2015,
  \apj, 805, 130

\bibitem[{{Koposov} {et~al.}(2018){Koposov}, {Walker}, {Belokurov}, {Casey},
  {Geringer-Sameth}, {Mackey}, {Da Costa}, {Erkal}, {Jethwa}, {Mateo},
  {Olszewski}, \& {Bailey}}]{2018MNRAS.479.5343K}
{Koposov}, S.~E., {Walker}, M.~G., {Belokurov}, V., {et~al.} 2018, \mnras, 479,
  5343

\bibitem[{{Koposov} {et~al.}(2022){Koposov}, {Erkal}, {Li}, {Da Costa},
  {Cullinane}, {Ji}, {Kuehn}, {Lewis}, {Pace}, {Shipp}, {Zucker},
  {Bland-Hawthorn}, {Lilleengen}, \& {Martell}}]{2022arXiv221104495K}
{Koposov}, S.~E., {Erkal}, D., {Li}, T.~S., {et~al.} 2022, arXiv e-prints,
  arXiv:2211.04495

\bibitem[{{Li} {et~al.}(2019){Li}, {Koposov}, {Zucker}, {Lewis}, {Kuehn},
  {Simpson}, {Ji}, {Shipp}, {Mao}, {Geha}, {Pace}, {Mackey}, {Allam}, {Tucker},
  {Da Costa}, {Erkal}, {Simon}, {Mould}, {Martell}, {Wan}, {De Silva},
  {Bechtol}, {Balbinot}, {Belokurov}, {Bland-Hawthorn}, {Casey}, {Cullinane},
  {Drlica-Wagner}, {Sharma}, {Vivas}, {Wechsler}, {Yanny}, \& {S5
  Collaboration}}]{2019MNRAS.490.3508L}
{Li}, T.~S., {Koposov}, S.~E., {Zucker}, D.~B., {et~al.} 2019, \mnras, 490,
  3508

\bibitem[{{Longeard} {et~al.}(2019){Longeard}, {Martin}, {Ibata}, {Collins},
  {Laevens}, {Bell}, \& {Mackey}}]{2019MNRAS.490.1498L}
{Longeard}, N., {Martin}, N., {Ibata}, R.~A., {et~al.} 2019, \mnras, 490, 1498

\bibitem[{{Longeard} {et~al.}(2018){Longeard}, {Martin}, {Starkenburg},
  {Ibata}, {Collins}, {Geha}, {Laevens}, {Rich}, {Aguado}, {Arentsen},
  {Carlberg}, {C{\^o}t{\'e}}, {Hill}, {Jablonka}, {Gonz{\'a}lez Hern{\'a}ndez},
  {Navarro}, {S{\'a}nchez-Janssen}, {Tolstoy}, {Venn}, \&
  {Youakim}}]{2018MNRAS.480.2609L}
{Longeard}, N., {Martin}, N., {Starkenburg}, E., {et~al.} 2018, \mnras, 480,
  2609

\bibitem[{{Luque} {et~al.}(2016){Luque}, {Queiroz}, {Santiago}, {Pieres},
  {Balbinot}, {Bechtol}, {Drlica-Wagner}, {Neto}, {da Costa}, {Maia}, {Yanny},
  {Abbott}, {Allam}, {Benoit-L{\'e}vy}, {Bertin}, {Brooks}, {Buckley-Geer},
  {Burke}, {Rosell}, {Kind}, {Carretero}, {Cunha}, {Desai}, {Diehl},
  {Dietrich}, {Eifler}, {Finley}, {Flaugher}, {Fosalba}, {Frieman}, {Gerdes},
  {Gruen}, {Gutierrez}, {Honscheid}, {James}, {Kuehn}, {Kuropatkin}, {Lahav},
  {Li}, {March}, {Marshall}, {Martini}, {Miquel}, {Neilsen}, {Nichol}, {Nord},
  {Ogando}, {Plazas}, {Romer}, {Roodman}, {Sanchez}, {Scarpine}, {Schubnell},
  {Sevilla-Noarbe}, {Smith}, {Soares-Santos}, {Sobreira}, {Suchyta}, {Swanson},
  {Tarle}, {Thaler}, {Tucker}, {Walker}, \& {Zhang}}]{2016MNRAS.458..603L}
{Luque}, E., {Queiroz}, A., {Santiago}, B., {et~al.} 2016, \mnras, 458, 603

\bibitem[{{Luque} {et~al.}(2018){Luque}, {Santiago}, {Pieres}, {Marshall},
  {Pace}, {Kron}, {Drlica-Wagner}, {Queiroz}, {Balbinot}, {dal Ponte}, {Fausti
  Neto}, {da Costa}, {Maia}, {Walker}, {Abdalla}, {Allam}, {Annis}, {Bechtol},
  {Benoit-L{\'e}vy}, {Bertin}, {Brooks}, {Carnero Rosell}, {Carrasco Kind},
  {Carretero}, {Crocce}, {Davis}, {Doel}, {Eifler}, {Flaugher},
  {Garc{\'\i}a-Bellido}, {Gerdes}, {Gruen}, {Gruendl}, {Gutierrez},
  {Honscheid}, {James}, {Kuehn}, {Kuropatkin}, {Miquel}, {Nichol}, {Plazas},
  {Sanchez}, {Scarpine}, {Schindler}, {Sevilla-Noarbe}, {Smith},
  {Soares-Santos}, {Sobreira}, {Suchyta}, {Tarle}, \&
  {Thomas}}]{2018MNRAS.478.2006L}
{Luque}, E., {Santiago}, B., {Pieres}, A., {et~al.} 2018, \mnras, 478, 2006

\bibitem[{{Mackey} \& {Gilmore}(2004)}]{2004MNRAS.352..153M}
{Mackey}, A.~D., \& {Gilmore}, G.~F. 2004, \mnras, 352, 153

\bibitem[{{Mackey} {et~al.}(2018){Mackey}, {Koposov}, {Da Costa}, {Belokurov},
  {Erkal}, \& {Kuzma}}]{2018ApJ...858L..21M}
{Mackey}, D., {Koposov}, S., {Da Costa}, G., {et~al.} 2018, \apjl, 858, L21

\bibitem[{{Martin} {et~al.}(2008){Martin}, {de Jong}, \& {Rix}}]{Martin:2008}
{Martin}, N.~F., {de Jong}, J.~T.~A., \& {Rix}, H.-W. 2008, \apj, 684, 1075

\bibitem[{{Martin} {et~al.}(2015){Martin}, {Nidever}, {Besla}, {Olsen},
  {Walker}, {Vivas}, {Gruendl}, {Kaleida}, {Mu{\~n}oz}, {Blum}, {Saha}, {Conn},
  {Bell}, {Chu}, {Cioni}, {de Boer}, {Gallart}, {Jin}, {Kunder}, {Majewski},
  {Martinez-Delgado}, {Monachesi}, {Monelli}, {Monteagudo}, {No{\"e}l},
  {Olszewski}, {Stringfellow}, {van der Marel}, \&
  {Zaritsky}}]{2015ApJ...804L...5M}
{Martin}, N.~F., {Nidever}, D.~L., {Besla}, G., {et~al.} 2015, \apjl, 804, L5

\bibitem[{{Martin} {et~al.}(2016){Martin}, {Jungbluth}, {Nidever}, {Bell},
  {Besla}, {Blum}, {Cioni}, {Conn}, {Kaleida}, {Gallart}, {Jin}, {Majewski},
  {Martinez-Delgado}, {Monachesi}, {Mu{\~n}oz}, {No{\"e}l}, {Olsen},
  {Stringfellow}, {van der Marel}, {Vivas}, {Walker}, \&
  {Zaritsky}}]{2016ApJ...830L..10M}
{Martin}, N.~F., {Jungbluth}, V., {Nidever}, D.~L., {et~al.} 2016, \apjl, 830,
  L10

\bibitem[{{Massana} {et~al.}(2020){Massana}, {No{\"e}l}, {Nidever}, {Erkal},
  {de Boer}, {Choi}, {Majewski}, {Olsen}, {Monachesi}, {Gallart}, {Marel},
  {Ruiz-Lara}, {Zaritsky}, {Martin}, {Mu{\~n}oz}, {Cioni}, {Bell}, {Bell},
  {Stringfellow}, {Belokurov}, {Monelli}, {Walker}, {Mart{\'\i}nez-Delgado},
  {Vivas}, \& {Conn}}]{2020MNRAS.498.1034M}
{Massana}, P., {No{\"e}l}, N. E.~D., {Nidever}, D.~L., {et~al.} 2020, \mnras,
  498, 1034

\bibitem[{{Massana} {et~al.}(2022){Massana}, {Ruiz-Lara}, {No{\"e}l},
  {Gallart}, {Nidever}, {Choi}, {Sakowska}, {Besla}, {Olsen}, {Monelli},
  {Dorta}, {Stringfellow}, {Cassisi}, {Bernard}, {Zaritsky}, {Cioni},
  {Monachesi}, {van der Marel}, {de Boer}, \& {Walker}}]{2022MNRAS.513L..40M}
{Massana}, P., {Ruiz-Lara}, T., {No{\"e}l}, N.~E.~D., {et~al.} 2022, \mnras,
  513, L40

\bibitem[{{Mau} {et~al.}(2019){Mau}, {Drlica-Wagner}, {Bechtol}, {Pace}, {Li},
  {Soares-Santos}, {Kuropatkin}, {Allam}, {Tucker}, {Santana-Silva}, {Yanny},
  {Jethwa}, {Palmese}, {Vivas}, {Burgad}, {Chen}, \& {BLISS
  Collaboration}}]{2019ApJ...875..154M}
{Mau}, S., {Drlica-Wagner}, A., {Bechtol}, K., {et~al.} 2019, \apj, 875, 154

\bibitem[{{Mau} {et~al.}(2020){Mau}, {Cerny}, {Pace}, {Choi}, {Drlica-Wagner},
  {Santana-Silva}, {Riley}, {Erkal}, {Stringfellow}, {Adam{\'o}w}, {Carlin},
  {Gruendl}, {Hernandez-Lang}, {Kuropatkin}, {Li}, {Mart{\'\i}nez-V{\'a}zquez},
  {Morganson}, {Mutlu-Pakdil}, {Neilsen}, {Nidever}, {Olsen}, {Sand},
  {Tollerud}, {Tucker}, {Yanny}, {Zenteno}, {Allam}, {Barkhouse}, {Bechtol},
  {Bell}, {Balaji}, {Crnojevi{\'c}}, {Esteves}, {Ferguson}, {Gallart},
  {Hughes}, {James}, {Jethwa}, {Johnson}, {Kuehn}, {Majewski}, {Mao},
  {Massana}, {McNanna}, {Monachesi}, {Nadler}, {No{\"e}l}, {Palmese},
  {Paz-Chinchon}, {Pieres}, {Sanchez}, {Shipp}, {Simon}, {Soares-Santos},
  {Tavangar}, {van der Marel}, {Vivas}, {Walker}, \&
  {Wechsler}}]{2020ApJ...890..136M}
{Mau}, S., {Cerny}, W., {Pace}, A.~B., {et~al.} 2020, \apj, 890, 136

\bibitem[{{Mazzi} {et~al.}(2021){Mazzi}, {Girardi}, {Zaggia}, {Pastorelli},
  {Rubele}, {Bressan}, {Cioni}, {Clementini}, {Cusano}, {Rocha}, {Gullieuszik},
  {Kerber}, {Marigo}, {Ripepi}, {Bekki}, {Bell}, {de Grijs}, {Groenewegen},
  {Ivanov}, {Oliveira}, {Sun}, \& {van Loon}}]{2021MNRAS.508..245M}
{Mazzi}, A., {Girardi}, L., {Zaggia}, S., {et~al.} 2021, \mnras, 508, 245

\bibitem[{{McConnachie}(2012)}]{2012AJ....144....4M}
{McConnachie}, A.~W. 2012, \aj, 144, 4

\bibitem[{{Moskowitz} \& {Walker}(2020)}]{2020ApJ...892...27M}
{Moskowitz}, A.~G., \& {Walker}, M.~G. 2020, \apj, 892, 27

\bibitem[{{Mu{\~n}oz} {et~al.}(2018){Mu{\~n}oz}, {C{\^o}t{\'e}}, {Santana},
  {Geha}, {Simon}, {Oyarz{\'u}n}, {Stetson}, \&
  {Djorgovski}}]{2018ApJ...860...66M}
{Mu{\~n}oz}, R.~R., {C{\^o}t{\'e}}, P., {Santana}, F.~A., {et~al.} 2018, \apj,
  860, 66

\bibitem[{{Mu{\~n}oz} {et~al.}(2012){Mu{\~n}oz}, {Geha}, {C{\^o}t{\'e}},
  {Vargas}, {Santana}, {Stetson}, {Simon}, \&
  {Djorgovski}}]{2012ApJ...753L..15M}
{Mu{\~n}oz}, R.~R., {Geha}, M., {C{\^o}t{\'e}}, P., {et~al.} 2012, \apjl, 753,
  L15

\bibitem[{{Mutlu-Pakdil} {et~al.}(2018){Mutlu-Pakdil}, {Sand}, {Carlin},
  {Spekkens}, {Caldwell}, {Crnojevi{\'c}}, {Hughes}, {Willman}, \&
  {Zaritsky}}]{2018ApJ...863...25M}
{Mutlu-Pakdil}, B., {Sand}, D.~J., {Carlin}, J.~L., {et~al.} 2018, \apj, 863,
  25

\bibitem[{{Nidever} {et~al.}(2008){Nidever}, {Majewski}, \& {Butler
  Burton}}]{2008ApJ...679..432N}
{Nidever}, D.~L., {Majewski}, S.~R., \& {Butler Burton}, W. 2008, \apj, 679,
  432

\bibitem[{{Nidever} {et~al.}(2017){Nidever}, {Olsen}, {Walker}, {Vivas},
  {Blum}, {Kaleida}, {Choi}, {Conn}, {Gruendl}, {Bell}, {Besla}, {Mu{\~n}oz},
  {Gallart}, {Martin}, {Olszewski}, {Saha}, {Monachesi}, {Monelli}, {de Boer},
  {Johnson}, {Zaritsky}, {Stringfellow}, {van der Marel}, {Cioni}, {Jin},
  {Majewski}, {Martinez-Delgado}, {Monteagudo}, {No{\"e}l}, {Bernard},
  {Kunder}, {Chu}, {Bell}, {Santana}, {Frechem}, {Medina}, {Parkash},
  {Navarrete}, \& {Hayes}}]{2017AJ....154..199N}
{Nidever}, D.~L., {Olsen}, K., {Walker}, A.~R., {et~al.} 2017, \aj, 154, 199

\bibitem[{{Pace} {et~al.}(2022){Pace}, {Erkal}, \& {Li}}]{2022ApJ...940..136P}
{Pace}, A.~B., {Erkal}, D., \& {Li}, T.~S. 2022, \apj, 940, 136

\bibitem[{{Patel} {et~al.}(2020){Patel}, {Kallivayalil}, {Garavito-Camargo},
  {Besla}, {Weisz}, {van der Marel}, {Boylan-Kolchin}, {Pawlowski}, \&
  {G{\'o}mez}}]{2020ApJ...893..121P}
{Patel}, E., {Kallivayalil}, N., {Garavito-Camargo}, N., {et~al.} 2020, \apj,
  893, 121

\bibitem[{{Piatti} \& {Lucchini}(2022)}]{2022MNRAS.515.4005P}
{Piatti}, A.~E., \& {Lucchini}, S. 2022, \mnras, 515, 4005

\bibitem[{{Pieres} {et~al.}(2017){Pieres}, {Santiago}, {Drlica-Wagner},
  {Bechtol}, {Marel}, {Besla}, {Martin}, {Belokurov}, {Gallart},
  {Martinez-Delgado}, {Marshall}, {N{\"o}el}, {Majewski}, {Cioni}, {Li},
  {Hartley}, {Luque}, {Conn}, {Walker}, {Balbinot}, {Stringfellow}, {Olsen},
  {Nidever}, {da Costa}, {Ogando}, {Maia}, {Neto}, {Abbott}, {Abdalla},
  {Allam}, {Annis}, {Benoit-L{\'e}vy}, {Rosell}, {Kind}, {Carretero}, {Cunha},
  {D'Andrea}, {Desai}, {Diehl}, {Doel}, {Flaugher}, {Fosalba},
  {Garc{\'\i}a-Bellido}, {Gruen}, {Gruendl}, {Gschwend}, {Gutierrez},
  {Honscheid}, {James}, {Kuehn}, {Kuropatkin}, {Menanteau}, {Miquel}, {Plazas},
  {Romer}, {Sako}, {Sanchez}, {Scarpine}, {Schubnell}, {Sevilla-Noarbe},
  {Smith}, {Soares-Santos}, {Sobreira}, {Suchyta}, {Swanson}, {Tarle},
  {Tucker}, \& {Wester}}]{2017MNRAS.468.1349P}
{Pieres}, A., {Santiago}, B.~X., {Drlica-Wagner}, A., {et~al.} 2017, \mnras,
  468, 1349

\bibitem[{Pieres {et~al.}(2017)Pieres, Santiago, Drlica-Wagner, Bechtol, {Van
  Der Marel}, Besla, Martin, Belokurov, Gallart, Martinez-Delgado, Marshall,
  N{\"o}el, Majewski, Cioni, Li, Hartley, Luque, Conn, Walker, Balbinot,
  Stringfellow, Olsen, Nidever, {Da Costa}, Ogando, Maia, {Fausti Neto},
  Abbott, Abdalla, Allam, Annis, Benoit-L{\'e}vy, {Carnero Rosell}, {Carrasco
  Kind}, Carretero, Cunha, D{\textquoteright}Andrea, Desai, Diehl, Doel,
  Flaugher, Fosalba, Garc{\'i}a-Bellido, Gruen, Gruendl, Gschwend, Gutierrez,
  Honscheid, James, Kuehn, Kuropatkin, Menanteau, Miquel, Plazas, Romer, Sako,
  Sanchez, Scarpine, Schubnell, Sevilla-Noarbe, Smith, Soares-Santos, Sobreira,
  Suchyta, Swanson, Tarle, Tucker, \& Wester}]{smcnod}
Pieres, A., Santiago, B., Drlica-Wagner, A., {et~al.} 2017, Monthly Notices of
  the Royal Astronomical Society, 468, 1349

\bibitem[{{Pietrzy{\'n}ski} {et~al.}(2013){Pietrzy{\'n}ski}, {Graczyk},
  {Gieren}, {Thompson}, {Pilecki}, {Udalski}, {Soszy{\'n}ski}, {Koz{\l}owski},
  {Konorski}, {Suchomska}, {Bono}, {Moroni}, {Villanova}, {Nardetto},
  {Bresolin}, {Kudritzki}, {Storm}, {Gallenne}, {Smolec}, {Minniti}, {Kubiak},
  {Szyma{\'n}ski}, {Poleski}, {Wyrzykowski}, {Ulaczyk}, {Pietrukowicz},
  {G{\'o}rski}, \& {Karczmarek}}]{2013Natur.495...76P}
{Pietrzy{\'n}ski}, G., {Graczyk}, D., {Gieren}, W., {et~al.} 2013, \nat, 495,
  76

\bibitem[{{Plummer}(1911)}]{1911MNRAS..71..460P}
{Plummer}, H.~C. 1911, \mnras, 71, 460

\bibitem[{{Richstein} {et~al.}(2022){Richstein}, {Patel}, {Kallivayalil},
  {Simon}, {Zivick}, {Tollerud}, {Fritz}, {Warfield}, {Besla}, {van der Marel},
  {Wetzel}, {Choi}, {Deason}, {Geha}, {Guhathakurta}, {Jeon}, {Kirby},
  {Libralato}, {Sacchi}, \& {Sohn}}]{2022ApJ...933..217R}
{Richstein}, H., {Patel}, E., {Kallivayalil}, N., {et~al.} 2022, \apj, 933, 217

\bibitem[{{Ripepi} {et~al.}(2014){Ripepi}, {Cignoni}, {Tosi}, {Marconi},
  {Musella}, {Grado}, {Limatola}, {Clementini}, {Brocato}, {Cantiello},
  {Capaccioli}, {Cappellaro}, {Cioni}, {Cusano}, {Dall'Ora}, {Gallagher},
  {Grebel}, {Nota}, {Palla}, {Romano}, {Raimondo}, {Sabbi}, {Getman},
  {Napolitano}, {Schipani}, \& {Zaggia}}]{2014MNRAS.442.1897R}
{Ripepi}, V., {Cignoni}, M., {Tosi}, M., {et~al.} 2014, \mnras, 442, 1897

\bibitem[{{Ripepi} {et~al.}(2017){Ripepi}, {Cioni}, {Moretti}, {Marconi},
  {Bekki}, {Clementini}, {de Grijs}, {Emerson}, {Groenewegen}, {Ivanov},
  {Molinaro}, {Muraveva}, {Oliveira}, {Piatti}, {Subramanian}, \& {van
  Loon}}]{2017MNRAS.472..808R}
{Ripepi}, V., {Cioni}, M.-R.~L., {Moretti}, M.~I., {et~al.} 2017, \mnras, 472,
  808

\bibitem[{{Ripepi} {et~al.}(2022){Ripepi}, {Chemin}, {Molinaro}, {Cioni},
  {Bekki}, {Clementini}, {de Grijs}, {De Somma}, {El Youssoufi}, {Girardi},
  {Groenewegen}, {Ivanov}, {Marconi}, {McMillan}, \& {van
  Loon}}]{2022MNRAS.512..563R}
{Ripepi}, V., {Chemin}, L., {Molinaro}, R., {et~al.} 2022, \mnras, 512, 563

\bibitem[{{Rubele} {et~al.}(2015){Rubele}, {Girardi}, {Kerber}, {Cioni},
  {Piatti}, {Zaggia}, {Bekki}, {Bressan}, {Clementini}, {de Grijs}, {Emerson},
  {Groenewegen}, {Ivanov}, {Marconi}, {Marigo}, {Moretti}, {Ripepi},
  {Subramanian}, {Tatton}, \& {van Loon}}]{2015MNRAS.449..639R}
{Rubele}, S., {Girardi}, L., {Kerber}, L., {et~al.} 2015, \mnras, 449, 639

\bibitem[{{Rubele} {et~al.}(2018){Rubele}, {Pastorelli}, {Girardi}, {Cioni},
  {Zaggia}, {Marigo}, {Bekki}, {Bressan}, {Clementini}, {de Grijs}, {Emerson},
  {Groenewegen}, {Ivanov}, {Muraveva}, {Nanni}, {Oliveira}, {Ripepi}, {Sun}, \&
  {van Loon}}]{2018MNRAS.478.5017R}
{Rubele}, S., {Pastorelli}, G., {Girardi}, L., {et~al.} 2018, \mnras, 478, 5017

\bibitem[{Sales {et~al.}(2016)Sales, Navarro, Kallivayalil, \&
  Frenk}]{10.1093/mnras/stw2816}
Sales, L.~V., Navarro, J.~F., Kallivayalil, N., \& Frenk, C.~S. 2016, Monthly
  Notices of the Royal Astronomical Society, 465, 1879.
\newblock \url{https://doi.org/10.1093/mnras/stw2816}

\bibitem[{{Schlafly} \& {Finkbeiner}(2011)}]{2011ApJ...737..103S}
{Schlafly}, E.~F., \& {Finkbeiner}, D.~P. 2011, \apj, 737, 103

\bibitem[{{Schlegel} {et~al.}(1998){Schlegel}, {Finkbeiner}, \&
  {Davis}}]{Schlegel:1998}
{Schlegel}, D.~J., {Finkbeiner}, D.~P., \& {Davis}, M. 1998, \apj, 500, 525

\bibitem[{{Simon} {et~al.}(2020){Simon}, {Li}, {Erkal}, {Pace},
  {Drlica-Wagner}, {James}, {Marshall}, {Bechtol}, {Hansen}, {Kuehn}, {Lidman},
  {Allam}, {Annis}, {Avila}, {Bertin}, {Brooks}, {Burke}, {Rosell}, {Carrasco
  Kind}, {Carretero}, {da Costa}, {De Vicente}, {Desai}, {Doel}, {Eifler},
  {Everett}, {Fosalba}, {Frieman}, {Garc{\'\i}a-Bellido}, {Gaztanaga},
  {Gerdes}, {Gruen}, {Gruendl}, {Gschwend}, {Gutierrez}, {Hollowood},
  {Honscheid}, {Krause}, {Kuropatkin}, {MacCrann}, {Maia}, {March}, {Miquel},
  {Palmese}, {Paz-Chinch{\'o}n}, {Plazas}, {Reil}, {Roodman}, {Sanchez},
  {Santiago}, {Scarpine}, {Schubnell}, {Serrano}, {Smith}, {Suchyta}, {Tarle},
  {Walker}, \& {DES Collaboration}}]{2020ApJ...892..137S}
{Simon}, J.~D., {Li}, T.~S., {Erkal}, D., {et~al.} 2020, \apj, 892, 137

\bibitem[{{Torrealba} {et~al.}(2019){Torrealba}, {Belokurov}, \&
  {Koposov}}]{2019MNRAS.484.2181T}
{Torrealba}, G., {Belokurov}, V., \& {Koposov}, S.~E. 2019, \mnras, 484, 2181

\bibitem[{{Torrealba} {et~al.}(2016{\natexlab{a}}){Torrealba}, {Koposov},
  {Belokurov}, \& {Irwin}}]{2016MNRAS.459.2370T}
{Torrealba}, G., {Koposov}, S.~E., {Belokurov}, V., \& {Irwin}, M.
  2016{\natexlab{a}}, \mnras, 459, 2370

\bibitem[{{Torrealba} {et~al.}(2016{\natexlab{b}}){Torrealba}, {Koposov},
  {Belokurov}, {Irwin}, {Collins}, {Spencer}, {Ibata}, {Mateo}, {Bonaca}, \&
  {Jethwa}}]{2016MNRAS.463..712T}
{Torrealba}, G., {Koposov}, S.~E., {Belokurov}, V., {et~al.}
  2016{\natexlab{b}}, \mnras, 463, 712

\bibitem[{{Torrealba} {et~al.}(2018){Torrealba}, {Belokurov}, {Koposov},
  {Bechtol}, {Drlica-Wagner}, {Olsen}, {Vivas}, {Yanny}, {Jethwa}, {Walker},
  {Li}, {Allam}, {Conn}, {Gallart}, {Gruendl}, {James}, {Johnson}, {Kuehn},
  {Kuropatkin}, {Martin}, {Martinez-Delgado}, {Nidever}, {No{\"e}l}, {Simon},
  {Stringfellow}, \& {Tucker}}]{2018MNRAS.475.5085T}
{Torrealba}, G., {Belokurov}, V., {Koposov}, S.~E., {et~al.} 2018, \mnras, 475,
  5085

\bibitem[{{van der Marel} \& {Kallivayalil}(2014)}]{2014ApJ...781..121V}
{van der Marel}, R.~P., \& {Kallivayalil}, N. 2014, \apj, 781, 121

\bibitem[{{van der Walt} {et~al.}(2011){van der Walt}, {Colbert}, \&
  {Varoquaux}}]{2011CSE....13b..22V}
{van der Walt}, S., {Colbert}, S.~C., \& {Varoquaux}, G. 2011, Computing in
  Science and Engineering, 13, 22

\bibitem[{{Virtanen} {et~al.}(2020){Virtanen}, {Gommers}, {Oliphant},
  {Haberland}, {Reddy}, {Cournapeau}, {Burovski}, {Peterson}, {Weckesser},
  {Bright}, {van der Walt}, {Brett}, {Wilson}, {Millman}, {Mayorov}, {Nelson},
  {Jones}, {Kern}, {Larson}, {Carey}, {Polat}, {Feng}, {Moore}, {VanderPlas},
  {Laxalde}, {Perktold}, {Cimrman}, {Henriksen}, {Quintero}, {Harris},
  {Archibald}, {Ribeiro}, {Pedregosa}, {van Mulbregt}, \& {SciPy 1. 0
  Contributors}}]{2020NatMe..17..261V}
{Virtanen}, P., {Gommers}, R., {Oliphant}, T.~E., {et~al.} 2020, Nature
  Methods, 17, 261

\bibitem[{{Wang} {et~al.}(2019){Wang}, {de Boer}, {Pieres}, {Li},
  {Drlica-Wagner}, {Koposov}, {Vivas}, {Pace}, {Santiago}, {Walker}, {Tucker},
  {Strigari}, {Marshall}, {Yanny}, {DePoy}, {Bechtol}, {Roodman}, {Abbott},
  {Abdalla}, {Allam}, {Annis}, {Avila}, {Bertin}, {Brooks}, {Burke}, {Carnero
  Rosell}, {Carrasco Kind}, {Cunha}, {D'Andrea}, {da Costa}, {De Vicente},
  {Desai}, {Eifler}, {Estrada}, {Flaugher}, {Frieman}, {Garc{\'\i}a-Bellido},
  {Gerdes}, {Gruen}, {Gruendl}, {Gutierrez}, {Hollowood}, {Honscheid}, {James},
  {Kuehn}, {Kuropatkin}, {Lahav}, {Maia}, {Miquel}, {Sanchez}, {Scarpine},
  {Sevilla-Noarbe}, {Smith}, {Smith}, {Sobreira}, {Suchyta}, {Swanson},
  {Tarle}, \& {DES Collaboration}}]{2019ApJ...881..118W}
{Wang}, M.~Y., {de Boer}, T., {Pieres}, A., {et~al.} 2019, \apj, 881, 118

\bibitem[{{Weisz} {et~al.}(2016){Weisz}, {Koposov}, {Dolphin}, {Belokurov},
  {Gieles}, {Mateo}, {Olszewski}, {Sills}, \& {Walker}}]{2016ApJ...822...32W}
{Weisz}, D.~R., {Koposov}, S.~E., {Dolphin}, A.~E., {et~al.} 2016, \apj, 822,
  32

\bibitem[{{Zonca} {et~al.}(2019){Zonca}, {Singer}, {Lenz}, {Reinecke},
  {Rosset}, {Hivon}, \& {Gorski}}]{2019JOSS....4.1298Z}
{Zonca}, A., {Singer}, L., {Lenz}, D., {et~al.} 2019, The Journal of Open
  Source Software, 4, 1298

\end{thebibliography}
\appendix
\section{Candidate DELVE~6 Members Identified in GAIA DR3}
\label{sec:pmtable}
In \tabref{PMs}, we summarize the properties of the three potential DELVE~6 member stars that we identified in \Gaia DR3. We refer the reader to \secref{prop} for a more detailed discussion of these sources. 

\begin{deluxetable}{l c c c c c c c c}[H]
\tablecolumns{8}
\tablewidth{0pt}
\tabletypesize{\small}
\tablecaption{\label{tab:PMs} Properties of Candidate Proper-Motion Members of DELVE~6}
\tablehead{\colhead{\Gaia Source ID}  &  \colhead{R.A. (deg)}  & \colhead{Dec. (deg)}  & \colhead{$\varpi$ (mas)} & \colhead{$\mu_{\alpha}\cos \delta$ (mas yr$^{-1}$)} & $\mu_{\rm \delta}$ (mas yr$^{-1}$) & $p_{\code{ugali}}$ & Comment}
\startdata
4698076296289954048 & 33.0582 & -66.0618 & $0.0 \pm 0.4$& 0.6 $\pm 0.5$ &  $-1.1 \pm 0.4$ & 1.00 & BHB Candidate\\ 
4698076296289956352 & 33.0771 & -66.0579 & 0.2 $\pm$ 0.6& $2.0 \pm 0.8$ & $-2.2 \pm 0.9$ & 0.99 & RGB star\\ 
4698076296289957248 & 33.0549 &  -66.0558 & $0.0 \pm 0.3$ &$-3.0 \pm 0.4$ &-12.2 $\pm 0.3$ & 0.00 & RHB? Non-member\\ 
\enddata
\tablecomments{The astrometric properties included above are all taken directly from Gaia DR3. Here, $p_{\code{ugali}}$ is the membership probability of each star (rounded to the nearest hundredth) as determined from our fiducial results, which favored the isochrone model with $\tau = 13.5 \Gyr$ and $\rm [Fe/H] = -2.19$ dex.}
\end{deluxetable}
\FloatBarrier 

\section{References for Literature Data Presented in Figure 4}
\label{sec:refs}
In \figref{PopComp}, we compare DELVE~6's absolute $V$-band magnitude ($M_V$) and azimuthally-averaged half-light radius ($r_{1/2}$) to the known population of MW satellite galaxies, globular clusters, and recently-discovered ultra-faint star clusters.  For the MW satellite galaxies, we adopted or derived measurements of $M_V$ and $r_{1/2}$ from 
\citet{2012AJ....144....4M}, \citet{2015ApJ...805..130K}, \citet{2015ApJ...813..109D}, \citet{2015ApJ...804L...5M}, \citet{2015ApJ...808L..39K}, \citet{2016ApJ...833L...5D}, \citet{2016ApJ...824L..14C}, \citet{2016MNRAS.459.2370T}, \citet{2016MNRAS.463..712T}, \citet{2017AJ....154..267C}, \citet{2018PASJ...70S..18H}, \citet{2019PASJ...71...94H},
 \citet{2018MNRAS.479.5343K},  \citet{2018ApJ...860...66M}, \citet{2018ApJ...863...25M}, \citet{2018MNRAS.475.5085T}, \citet{2019ApJ...881..118W},  \citet{2020ApJ...892...27M},   \citet{2020ApJ...892..137S}, \citet{2020ApJ...890..136M}, \citet{2021ApJ...921...32J}, \citet{2021ApJ...920L..44C}, \cite{2021ApJ...916...81C}, \citet{2022arXiv220912422C}, \citet{2022ApJ...933..217R}, and \citet{2023ApJ...942..111C}. For the ultra-faint star clusters, we adopted or derived measurements from
\citet{2011AJ....142...88F}, \citet{2012ApJ...753L..15M}, \citet{2013ApJ...767..101B},  \citet{2015ApJ...799...73K}, \citet{2015ApJ...803...63K}, \citet{2016ApJ...822...32W},  \citet{2016ApJ...830L..10M},  \citet{2016ApJ...820..119K}, \citet{2018MNRAS.480.2609L}, \citet{2018MNRAS.478.2006L},  
\citet{2018ApJ...860...66M}, \citet{2018ApJ...863...25M},
\citet{2018ApJ...852...68C}, \citet{2019MNRAS.484.2181T}, \citet{2019MNRAS.490.1498L}, \citet{2019PASJ...71...94H}, \citet{2019ApJ...875..154M}, \citet{2020ApJ...890..136M}, \citet{2021ApJ...910...18C}, \citet{2022arXiv220912422C}, \citet{2022ApJ...929L..21G}. In cases where the above studies of ultra-faint systems quote only an elliptical half-light radius, we have converted these measurements to azimuthally-averaged half-light radii using these works' reported ellipticities (where possible). Lastly, measurements for the classical globular clusters are taken from \citet{2010arXiv1012.3224H}. Two ultra-faint star clusters discovered in the Sloan Digital Sky Survey (Koposov 1 and Koposov 2; \citealt{2007ApJ...669..337K}) have been removed from this catalog to avoid duplication.

\end{document}